\documentclass[aps,prd,twocolumn,amsmath,amssymb,showpacs,floatfix,nofootinbib]{revtex4}

\usepackage{amsmath,amsfonts,amssymb,slashed}
\usepackage{bbold,wasysym}

\usepackage[usenames]{xcolor} 

\usepackage{soul}
\usepackage{graphicx}
\usepackage{color}

\DeclareMathAlphabet\mathbfcal{OMS}{cmsy}{b}{n}
\definecolor{darkgreen}{cmyk}{0.85,0.2,1.00,0.35} 
\definecolor{purple}{cmyk}{0.5,1.0,0,0} 
\definecolor{darkblue}{cmyk}{1.0,1.0,0,0}
\usepackage{hyperref}
\hypersetup{
   colorlinks = true,
   allcolors = darkblue}
\definecolor{RedWine}{rgb}{0.743,0,0}
\definecolor{RoyalBlue}{rgb}{0.25,.41,.88}

\setstcolor{Blue}\usepackage{tabularx,ragged2e,booktabs}

\definecolor{darkgreen}{RGB}{0,80,40}

\newcolumntype{C}[1]{>{\Centering}m{#1}}

\renewcommand{\sec}{\ensuremath{\mathrm{s}}}
\newcommand{\mnras}{Mon. Not. R. Astron. Soc.}

\newcommand{\km}{\ensuremath{\mathrm{km}}}
\newcommand{\Mpc}{\ensuremath{\mathrm{Mpc}}}

\newcommand{\eV}{\ensuremath{\mathrm{eV}}}

\newcommand{\LambdaCDM}{\ensuremath {\Lambda {\rm CDM}}}

\begin{document}

\title{Cosmological tests of an axiverse-inspired
     quintessence field}
\author{Razieh Emami$^{1,2}$}\email{iasraziehm@ust.hk}
\author{Daniel Grin$^{3}$}\email{dgrin@kicp.uchicago.edu}
\author{Josef Pradler$^{4}$}\email{josef.pradler@oeaw.ac.at}
\author{Alvise Raccanelli$^{2}$}\email{alvise@jhu.edu}
\author{Marc Kamionkowski$^{2}$}\email{kamion@jhu.edu}

\affiliation{$^{1}$Institute for Advanced Study, Hong Kong University of Science and Technology, Hong Kong}
\affiliation{$^{2}$Department of Physics and Astronomy, Johns Hopkins University, 3400 N. Charles Street, Baltimore, Maryland 21218, USA}
\affiliation{$^{3}$Kavli Institute for Cosmological Physics and Department of Astronomy \& Astrophysics, University of Chicago, Chicago, IL, 60637, USA}
\affiliation{$^{4}$Institute of High Energy Physics, Austrian Academy of Sciences, Nikolsdorfergasse 18, 1050 Vienna, Austria}

\begin{abstract}
 Inspired by the string axiverse idea, it has been suggested that the recent transition from decelerated to accelerated cosmic expansion is
  driven by an axion-like quintessence field with a sub-Planckian
  decay constant. The
  scenario requires that the axion field be rather near the
  maximum of its potential, but is less finely tuned than other explanations of cosmic acceleration.  The model is parametrized by an axion
  decay constant $f$, the axion mass $m$,  and an initial
  misalignment angle $|\theta_i|$ which is close to
  $\pi$.  In order to determine the $m$ and $\theta_{i}$ values consistent with observations, these parameters are mapped onto observables: the Hubble parameter $H(z)$ at an angular diameter distance $d_{A}(z)$ to redshift $z= 0.57$, as well as the angular sound horizon of the cosmic microwave background (CMB). Measurements of the baryon acoustic oscillation (BAO) scale at $z\simeq 0.57$ by the BOSS survey and {Planck} measurements of CMB temperature anisotropies are then used to probe the $\left\{m,f,\theta_i\right\}$ parameter space. With current data, CMB constraints are the most powerful, allowing a fraction of only $\sim 0.2$ of the parameter-space volume.  Measurements of the BAO scale made using the SPHEREx or SKA experiments could go further, observationally distinguishing all but $\sim 10^{-2}$ or $\sim 10^{-5}$ of the parameter-space volume (allowed by simple priors) from the $\Lambda$CDM model.  \end{abstract}


\maketitle
\date{\today}

\section{Introduction}
\label{sec:intro}

The cause of the accelerated cosmic expansion \cite{Riess:1998cb,Perlmutter:1998np}
remains elusive. One possibility is quintessence, in which the acceleration is driven by the potential energy of a scalar field displaced from the 
minimum of its potential. The idea that late-time acceleration could be connected to a new fundamental scalar was suggested in Refs.~\cite{1977GReGr...8..833E,1982GReGr..14..769E,Fujii:1982ms,dolgovbook1983,zeebook1985,Nilles:1985kn,Bertolami:1986uj,1986PhLA..113..467S,Ford:1987de,Weiss:1987xa,Wetterich:1987fk,Wetterich:1987fm,Barr:1988zk,Singh:1988pp,Fujii:1989qk,Barrow:1990nv,1990AN....311..159A} (see Ref. \cite{Sahni:1999gb,Peebles:2002gy,Caldwell:2009ix} for a comprehensive review) and developed into viable models in Refs.~\cite{Ratra:1987rm,Caldwell:1997ii,Wetterich:1994bg,Coble:1996te,Turner:1998ex}.  
Alternative gravity solutions to cosmic acceleration or models
based upon large extra dimensions also generally require some
parameter to be tuned to be extremely small 
\cite{Copeland:2006wr,Caldwell:2009ix,Silvestri:2009hh}.  String landscape \cite{Bousso:2000xa} and/or
anthropic arguments \cite{Weinberg:1987dv} explain the small
value of the cosmological constant by supposing a Universe with
a cosmological constant of this value just happens to be the
one, of $\sim10^{120}$, that allows intelligent observers.

Quintessence models do not generically address the ``why now?''
problem; i.e., why the Universe transitions from decelerated
expansion to accelerated expansion only fairly recently, after
the Universe has cooled 30 orders of magnitude below the Planck
temperature.  There are so called ``tracker models," \cite{Peebles:1987ek,Ratra:1987rm,Caldwell:1997ii} in which the functional form of the quintessence potential causes its energy density to scale as the dominant component until late times, when its equation-of-state parameter $w\simeq -1$ and it comes to dominate the cosmic expansion and drive late-time acceleration. These models obviate the need for finely tuned initial conditions but still require a finely tuned overall energy scale for quintessence \cite{Peebles:1987ek,Ratra:1987rm,Caldwell:1997ii}.

Other quintessence models typically require a potential that is
extremely flat.  This latter problem may be
solved~\cite{Frieman:1995pm,Carroll:1998zi} if the quintessence field
is an axion-like field, in which case the shift symmetry of the
axion potential protects the extraordinary flatness required of
the quintessence potential.  

This solution, though, still
requires either that the axion decay constant have a
super-Planckian value, something that violates the
gravity-as-the-weakest force conjecture
\cite{ArkaniHamed:2006dz}, or that the initial
axion misalignment angle is extremely close to the value $\pi$ that 
maximizes the potential.  This latter option  is considered
unappealing as it replaces the fine-tuning of a cosmological
constant with the fine tuning of some other parameter.  The
hypothesis that quintessence is an axion-like field also
requires that the dark-energy density today must still be put in
by hand.

In recent work \cite{Kamionkowski:2014zda}, several of us
suggested that some of the remaining problems with axion-like
quintessence may be addressed in the string axiverse
\cite{Svrcek:2006yi,Arvanitaki:2009fg}.  String theories may
lead to a family of $O(100)$ axion-like fields with masses that
span a huge number of decades \cite{Svrcek:2006yi}.  In Ref.~\cite{Kamionkowski:2014zda}, a scenario was proposed in which these axions have masses that are
distributed roughly evenly per decade of axion mass. In such models, ultralight axions can furnish dark matter \textit{and} dark energy candidates. The theoretical motivation for and phenomenology of ultralight axions is reviewed at length in Ref. \cite{Marsh:2015xka}. 

In this scenario, in each decade of the cosmic expansion, one of
these axions becomes dynamical and has some small chance to drive an
accelerated expansion.  This chance is determined by the
initial value, assumed to be selected at random, of the axion
misalignment angle.  It was shown that with reasonable parameter
selections, there is a one in $\sim$100 chance that the Universe
will expand by $\sim30$ decades in scale factor before it
undergoes accelerated expansion.  The scenario thus explains
quintessence in terms of an axion-like field with dimensionless
parameters that never differ from unity by more than an order of
magnitude and also addresses the ``why now?'' problem.

While the scenario may lead to a variety of observational
consequences, the most basic prediction is that dark energy is
an axion-like quintessence field with a decay constant very
close to, but just below, the Planck mass.  The purpose of this
paper is to explore the observational consequences of this type
of quintessence theory, to impose constraints to the
parameter space from current measurements, and to forecast the
sensitivity of future measurements to the remaining regions of
the model parameter space. It is found that current cosmic microwave background measurements reduce the allowed parameter space to a fraction $\sim 0.2$ of the prior volume, while future measurements of the baryon acoustic oscillation scale by galaxy surveys could distinguish all but a fraction of $\sim 10^{-5}$ of the parameter-space volume (allowed by simple priors) from the $\Lambda$CDM model. On the other hand, it is found that if constraints are projected onto the two-dimensional space of axion mass $m$ and initial misalignment angle $\theta_{i}$, the sensitivity of future experiments improves only modestly upon current constraints,
assuming that the correct underlying dark-energy model is in fact the fiducial $\Lambda$CDM scenario.

We begin in Sec.~\ref{sec:model} by introducing the axiverse inspired quintessence model that is the focus of this paper. In Sec.~\ref{sec:dynamics} we obtain the relevant equations of motion, and discuss the resulting expansion histories. In Sec.~\ref{sec:conditions}, we identify axion model parameter choices that give appropriate values of the dark energy density today. In Sec.~\ref{sec:predictions}, we explore the probability distribution of dark energy density $\Omega_{\rm DE}$ and present-day equation-of-state parameter $w$
under flat priors for the axion mass $m$ (in units of $H_{0}$) and initial axion displacement $\delta=\pi-\theta_{i}$. In Sec.~\ref{sec:currentconstraints}, we obtain constraints on $m$ and $\delta$ for different values of the dimensionless Peccei-Quinn energy scale $\alpha=f/M_{\rm pl}$ using measurements of the cosmic expansion history from the baryon acoustic oscillation scale and the CMB. We conclude in Sec.~\ref{sec:conclusions}. In the Appendix, we discuss in detail our treatment of CMB anisotropies in axiverse-inspired quintessence models.

\section{Model}
\label{sec:model}

The scenario we consider is one in which cosmic acceleration is
due to the slow rolling of a quintessence field $\phi$ down
toward the minimum of its potential $V(\phi)$.  The standard
axion-like potential is
\begin{equation}
     V(\phi) = \Lambda^4 \left[ 1 - \cos(\phi/f)\right],
\label{eqn:standardpotential}
\end{equation}
where $f$ is the axion decay constant, and $\Lambda^4$ is a
vacuum-energy density comparable to the dark-energy density
today.  Below we refer to this standard axion model as Model A. In the language of Ref. \cite{Caldwell:2005tm}, this is a ``thawing" potential for dark energy \cite{Marsh:2014xoa}. As noted in 
Ref.~\cite{Kamionkowski:2014zda}, in this scenario, the many axions with mass $m \gtrsim 10^{-17}~{\rm eV}$ (which do not contribute to dark energy today) might dominate the energy density of the universe, overclosing it in the process. One way around this problem is to invoke the decay of such axions into standard-model particles. Another is to consider a different potential:
\begin{equation}
     V(\phi) = \Lambda^4 \left[ 1 - \cos(\phi/f)\right]^3.
\label{eqn:alteredpotential}
\end{equation}
We refer to this potential as Model B; it has the added advantage that after beginning to coherently oscillate, higher mass axions become energetically subdominant, leaving the standard cosmological expansion history in place (up to the time evolution of lighter axion fields, which contribute to the dark energy). 
In either case, the field $\phi$ takes on values $-\pi \leq
(\phi/f)\leq \pi$.  We also define the axion mass to be $m =
\Lambda^2/f$, although the field in the latter potential is in
fact massless.  We define the misalignment angle to be $\theta
\equiv \phi/f$ and often use it as a proxy for $\phi$.  The
model is also specified by an initial misalignment angle
$|\theta_i|$ which, as we will see, must be very close to
$\pi$.  We thus define $\delta_{i} \equiv \pi-|\theta_i|$.  The
time derivative of $\theta$ is assumed to be zero at time
$t=0$.  For reasons discussed in
Ref.~\cite{Kamionkowski:2014zda} (and as seen below),
the decay constant can be written as $f=\alpha M_p$, where $M_p=
( 8 \pi G)^{-1/2}=2.43\times 10^{18}$~GeV is the reduced Planck
mass, with values $\alpha \sim 0.1-1$.

In summary, the model is parametrized (for either potential) by
three quantities, which we take to be (1) the axion mass
parameter $m$, (2) the decay-constant $\alpha$, and
(3) the initial displacement $\delta_{i}$ of the misalignment
angle.
\section{Dynamics}
\label{sec:dynamics}
\begin{figure*}[tb]
\centering
\includegraphics[width=0.9\textwidth]{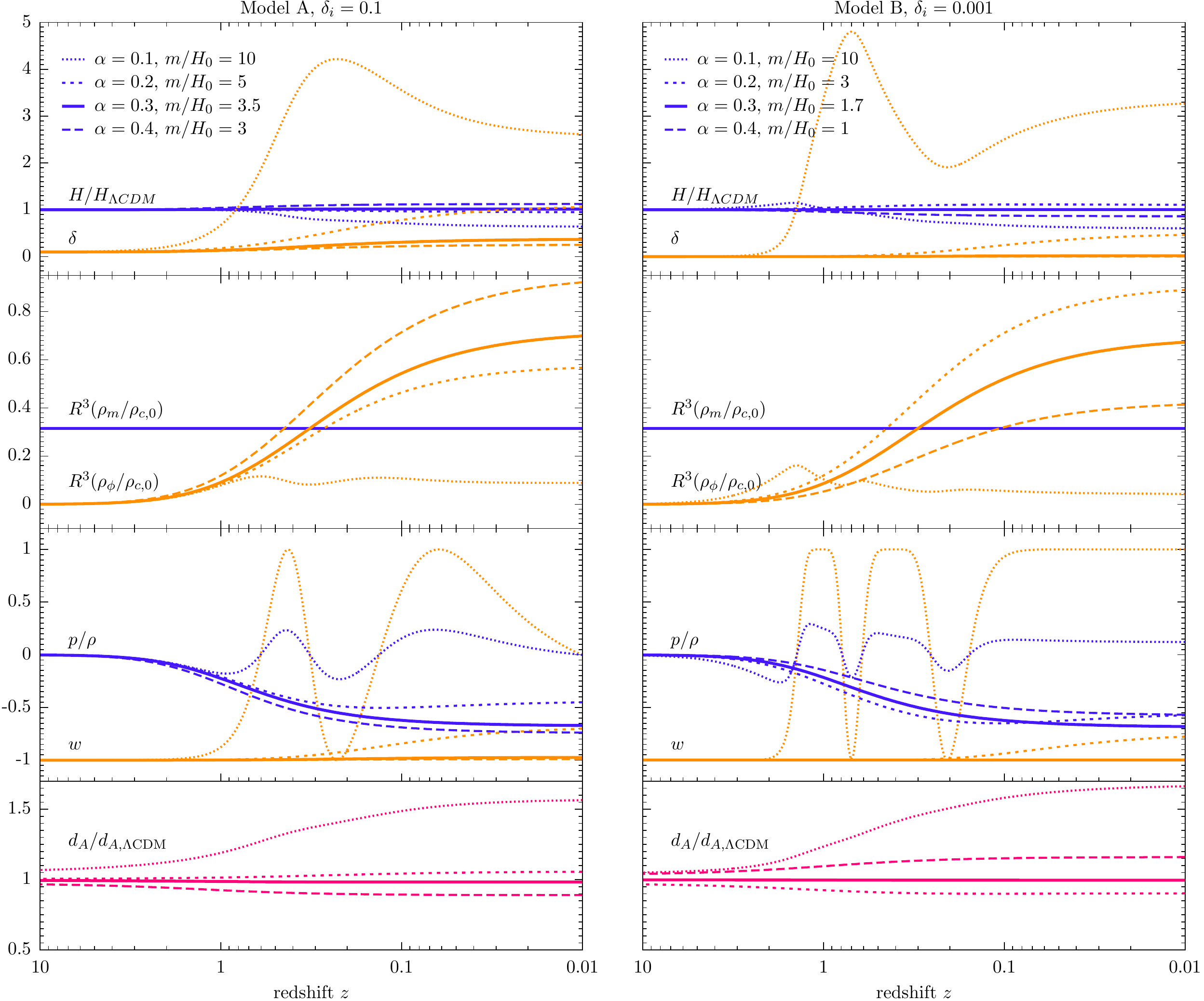}
\caption{Evolution of axion fields and associated
  cosmological parameters as a function of redshift $z$ for Model~A
  (left panel) and Model~B (right panel) for the model parameters as
  labeled. The top row shows the evolution of $\delta=\pi-\phi/f$ and the Hubble
  rate in comparison to \LambdaCDM. The
  second row shows the matter and axion density parameters, measured
  in units of $\rho_{c,0} = 3 H_0^2 M_P^2$. The third row shows the
  equation-of-state parameter $w$ of the axion, Eq.~(\ref{eq:eos}),
  and the overall ratio of total pressure to total energy density of
  all components. Finally, the bottom row is the ratio of the angular
  diameter distance for the axion model to the \LambdaCDM\ one.  The
  scenarios depicted by the thick line are consistent with observational constraints.}
\label{fig:evol}
\end{figure*}

The evolution of the field is determined by its equation of
motion,
\begin{equation}
     \ddot\phi + 3 H \dot \phi +V'(\phi)=0,
\label{eqn:EOM}
\end{equation}
where the dot denotes a derivative with respect to time and the prime
a derivative with respect to $\phi$. The Hubble rate $H$ is
given by the Friedmann equation,
\begin{equation}
     H = \frac{\dot a}{a} =   \frac{1}{\sqrt{3}
     M_p} \left(\rho_{\phi} + \rho_m +\rho_r +\rho_{\nu}^{m}
     \right)^{1/2} ,\label{eqn:hubble}
\end{equation}where $a$ is the cosmological scale factor.
Here, $\rho_{m(r)}$ is the matter (radiation) density and the energy
density of the axion is given by, 
\begin{align}
  \rho_{\phi} = \frac12 \dot\phi^2 +V(\phi) . 
\end{align}
The pressure $p_{\phi}$ is given by the same equation but an alternate
sign in front $V$. The equation-of-state parameter for this
quintessence field is given by,
\begin{equation}
\label{eq:eos}
     w = p_{\phi}/ \rho_{\phi}. 
\end{equation}
The dependence with redshift $z$ is
$\rho_m(z) = \Omega_{m} \rho_{c,0}(1+z)^3$ and
$\rho_r(z) = \Omega_{r}\rho_{c,0} (1+z)^4$ where $z$ is related to the scale
factor by $a=1/(1+z)$. The radiation energy density $\rho_{r}(z)$ includes the contribution of photons and massless neutrinos, while $\rho_{\nu}^{m}(z)$ is the energy density of massive neutrinos. The dependence of their energy density with redshift transitions from photon-like to matter-like when the temperature $T$ is comparable to the neutrino mass $m_{\nu}$ with numerical solutions obtained using standard expressions for a Fermi-Dirac distribution (e.g. Refs. \cite{Shaw:2009nf,cambnotes}).

It is then convenient to introduce the density parameters
$\Omega_i \equiv \rho_i/\rho_c$ where $\rho_c$ is the critical energy
density with present-day value $\rho_{c,0} = 3 H_0^2 M_P^2 $ where
$H_0 = 100\, h\, \km/\sec/\Mpc$, and we assume the best-fit Planck $\Lambda$CDM values for the dimensionless Hubble constant $h=0.67$ \cite{Ade:2015xua}.

Assuming that the dark matter density and the energy density in the
quintessence field are unrelated quantities, we fix the matter content
to the CMB-inferred physical baryon and cold dark matter
densities using the ``Planck TT + lowP'' values of Table~3
of~Ref.~\cite{Ade:2015xua}, $\Omega_b h^2 = 0.02222$ and 
$\Omega_{\rm cdm} h^2 = 0.1197$, respectively;
$\Omega_m = \Omega_b + \Omega_{\rm cdm}$.%
\footnote{In our numerical solutions, we
assume $2.04$ massless and one massive neutrino with $m_{\nu,3} = 0.06\,\eV$, for consistency with the choices made in Ref. \cite{Ade:2015xua}.}

Finally, we further assume a flat Universe and a
present day photon temperature
$T_0 = T(z=0) = 2.726$~K~\cite{Ade:2015xua}.

The equations of motion~(\ref{eqn:EOM}) together with the scale factor
$a$ are integrated numerically. We show the scalar field value $\phi$ using its instantaneous displacement from its maximum-energy value, given by the variable $\delta\equiv \pi-\phi/f$. We use a high initial redshift
$z_s = {\rm few}\times 1000$ whose precise value is numerically
inconsequential, but sufficiently large that the quintessence
energy density is initially negligible and $\dot\phi(z_s) = 0$ can be assumed.
Figure~\ref{fig:evol} shows some representative solutions for Model~A (left
panel) and Model~B (right panel) as a function of redshift $z$ for the
model parameters as labeled. The top row gives the evolution of
$\delta$ and the Hubble rate in comparison to the \LambdaCDM\
case. Integration stops when $a=1$. The second row shows the matter and
axion density parameters, and the third row shows the
equation-of-state parameter $w$ as well as the overall ratio of total
pressure to total energy density of all components. 

Finally, the angular diameter distance (for a flat Universe) is given
by,
\begin{align}
  d_A(z) = (1+z)^{-1}  \int_{t(z)}^{t_0} \frac{dt}{a[t(z)]} , 
\end{align}
where $t_0$ is the cosmic time today (at redshift $z=0$). The lines in
the bottom row of Fig.~\ref{fig:evol} show $d_A$ normalized to the
angular diameter distance of a flat $\LambdaCDM$ Universe with
constant $\Omega_{\Lambda} = 0.685$~\cite{Agashe:2014kda}.

The scenarios depicted by the thick line are visually indistinguishable from $\Lambda$CDM, and as we see in Sec.~\ref{sec:currentconstraints} and the Appendix, are consistent with observational constraints. At the $\delta$ values shown, for lower $\alpha$ values or higher $m$ values, field evolution is fast enough to distinguish by eye from $\Lambda$CDM. The deviation of $H(z)$ and $d_{A}(z)$ hint that observations sensitive to the cosmological expansion rate and specific angular-diameter distances (like measurements of baryon acoustic oscillations and CMB anisotropies) will test the axiverse-inspired quintessence model. For axion masses $m\gtrsim 10 H_{0}$, the field rolls quickly enough to enter the oscillatory regime (unless $\delta$ is finely tuned), spoiling the utility of the axion as a dark-energy candidate, as we see more quantitatively in Sec.~ \ref{sec:currentconstraints}.

\section{Conditions for dark energy}
\label{sec:conditions}

The first issue we address is the range of values of $\delta_i$ that
give rise to dark energy domination today. We begin with a rough analytic argument.
In Ref.~\cite{Kamionkowski:2014zda}, it was argued that $w\lesssim -1/3$ is
achieved if the slow-roll condition
$\epsilon = (M_p^2/2)(V'/V)^2 \lesssim 1$ is satisfied. Writing
$V(\theta)=\Lambda^4U(\theta)$, for $U(\theta)=1-\cos\theta$, this
suggests that $w\leq-1/3$ will be achieved if
$\delta = \pi-|\theta| \lesssim 2\sqrt{2}\alpha$, or
$\delta\lesssim \alpha$; for $U(\theta) = (1-\cos\theta)^3$, this is
amended to $\delta \lesssim \alpha/3$.  However, evolution of the
field from the intial time $z\gg1$ to the redshifts $z\sim1$ at which
dark energy domination begins implies that the value of $\delta$ at
$z\sim1$ is not necessarily the same as the initial value $\delta_i$
at $z\gg 1$. 

This evolution becomes significant for $m\gtrsim H_0$, and thus indicates a
quintessence field that begins to roll before the present day.
\begin{table}[tb]
\label{tab:one}
\caption{The range of values of the maximal initial misalignment
  angle $\delta_i^{\rm max}$ and axion mass $m$ consistent,
  for different values of the decay-constant $\alpha$, with
  a current dark energy equation-of-state parameter $-1\leq w\leq-0.7$ and
  dark energy density $ 0.6 \leq \Omega_{\phi} \leq 0.7$.
}

\begin{ruledtabular}
\begin{tabular}{ccccc}
 & \multicolumn{2}{c}{Model A}  & \multicolumn{2}{c}{Model B} \\
  \cline{2-3}  \cline{4-5} 
 $\alpha$ & $\delta_i^{\rm max}$ & $m/H_0$  & $\delta_i^{\rm max}$ & $m/H_0$\\ 
\colrule 
0.1 & $ 3.5\times10^{-3}$ & 9.5--10.5  & $5\times10^{-6}$ & 4.7--5.3  \\     
0.2 & $ 0.12$ & 4.7--5.5              & $ 5\times 10^{-3}$ & 2.38--2.68  \\ 
0.3 & $ 0.4$ & 3.1--3.9               & $ 0.04$ & 1.57--1.83  \\            
0.4 & $ 0.65$ & 2.4--3.2              & $ 0.12$ & 1.20--1.30  \\            
\end{tabular}
\end{ruledtabular}
\end{table}

These arguments can be made more quantitative. We numerically evolve Eqs.~(\ref{eqn:EOM}) and ~(\ref{eqn:hubble}), using the values $\alpha=0.1$, $0.2$, $0.3$, $0.4$. We determine a range of values for the dimensionless axion mass $m/H_{0}$ by imposing the condition $0.6\leq\Omega_{\phi}\leq0.7$ under the no-roll approximation. We then use the numerical solutions to determine the maximum initial value $\delta_{i}^{\rm max}$ of the field displacement needed to satisfy the constraint $-1\leq w\leq-0.7$. The results are shown in Table I.  The
results are relevant for the string axiverse scenario described in
Ref.~\cite{Kamionkowski:2014zda}.  There it was argued that the
fraction of the $-\pi\leq \theta_i \leq\pi$ range for the initial
misalignment angle that would give rise to cosmic acceleration was
$\sim \alpha$.  This {\it ansatz} then led, through further rough
calculations, to an estimate for the $\sim0.001-0.01$ probability
that, for $\alpha\sim 0.1-1$, a given Universe would have a
dark-energy density like that we observe.  That estimate, though,
neglected the early evolution of the scalar field.  We now see, with
our more precise calculation, that the real probability drops several
orders of magnitude below $\sim\alpha$ for $\alpha \lesssim 0.1$ (for
Model A) and for $\alpha \lesssim 0.2$ (for Model B).  We thus
conclude that the string-axiverse solution to the ``why now'' problem
described in Ref.~\cite{Kamionkowski:2014zda} requires
$\alpha \gtrsim 0.1$ for Model~A and $\alpha \gtrsim 0.2$ for Model~B.

Our numerical results for $0.1 \lesssim \alpha \lesssim 0.4$ also
suggest that for these allowable ranges of $\alpha$, the probability
that a given Universe has a dark-energy density like that we observe (parameterized by the allowed range of $\delta$)
is reduced by roughly an order of magnitude relative to the estimates
given in Ref.~\cite{Kamionkowski:2014zda}. Of course the conditions $0.6\leq\Omega_{\phi}\leq0.7$ and $-1\leq w\leq-0.7$ (which describe the properties of dark energy today) only crudely approximate the precise observational constraints to dark energy from actual data (like the galaxy correlation function and CMB power spectra), which are more precisely evaluated in Sec. \ref{sec:currentconstraints} and Appendix \ref{sec:cmb}. For now, we move on to make predictions for cosmological observables using the parameter-space regions identified in Table I.

\section{Predictions for quintessence observables}
\label{sec:predictions}

Suppose now that dark energy is indeed a quintessence field of the
form we consider here.  The string axiverse motivation requires, as we
just learned, $\alpha \gtrsim 0.1$.  An obvious question is what histories $w(z)$, $H(z)$, and $d_A(z)$ are predicted by the parameter values roughly consistent with this scenario for quintessence, especially in light of the fact that a whole suite of recent/ongoing
cosmological measurements~\cite{Albrecht:2006um,Linder:2008pp,Frieman:2008sn,Weinberg:2012es} target observables related to the recent history of these variables. We now address this question.

In order to address this question, we fix a value of $\alpha$ and then
assume that models are distributed uniformly in axion mass $m$ and
uniformly in the initial misalignment angle $\delta_i$, \textit{i.e.},
the joint probability distribution of models in the $m$-$\delta_i$
space is $\partial^2N/\partial m \partial \delta_i =$constant. 
In certain regions of parameter space there exists a one-to-one
mapping of the $(m,\delta_i)$ parameter space to
$(w, \Omega_{\rm de})$, where $\Omega_{\rm de}\equiv \Omega_{\phi}$. 

The distribution of models in the $(w,\Omega_{\rm de})$ parameter
space is then given by the Jacobian of the
transformation,~\textit{i.e.},
$\partial^2N/ \partial w \partial\Omega_{\rm de} \propto \partial
(m,\delta_i)/\partial(w,\Omega_{\rm de})$,
which we obtain numerically.  The probability distribution function
(PDF) $p(w)$ that arises for $w$, given a family of models uniformly
distributed in $m$ and $\delta_i$ is then,
\begin{equation}
     p(w) = \int d\Omega_{\rm de}\frac{1}{N}
     \frac{\partial^2N}{\partial\Omega_{\rm de} \partial w} \equiv
     \frac{1}{N} \frac{d N}{dw},
\label{eqn:wdistribution}
\end{equation}
where we begin from a number of $N$ prospective dark energy solutions
$(\delta_i, m)$ that satisfy the restrictions $0.6\leq\Omega_{\rm DE}\leq0.7$ and $w(z=0)\leq-0.7$
following Sec.~\ref{sec:conditions}. The expression for
$ p(\Omega_{\rm de})$ is obtained by the replacement
$ \Omega_{\rm de} \leftrightarrow w $ in~(\ref{eqn:wdistribution}).
We also wish to infer the PDF of other quantities, such as distance
measures $d_A$ at $z=1$ or $z_{\rm dec}\simeq 1090$ (where $z_{\rm dec}$ is the redshift of
decoupling). Starting from the same set of generated dark energy
solutions, the PDF of an observable $y(\delta_i,m)$ is numerically
straightforward to obtain by generalizing Eq.~(\ref{eqn:wdistribution}).
\begin{figure*}[h!tb]
\centering
\includegraphics[width=6.1 in]{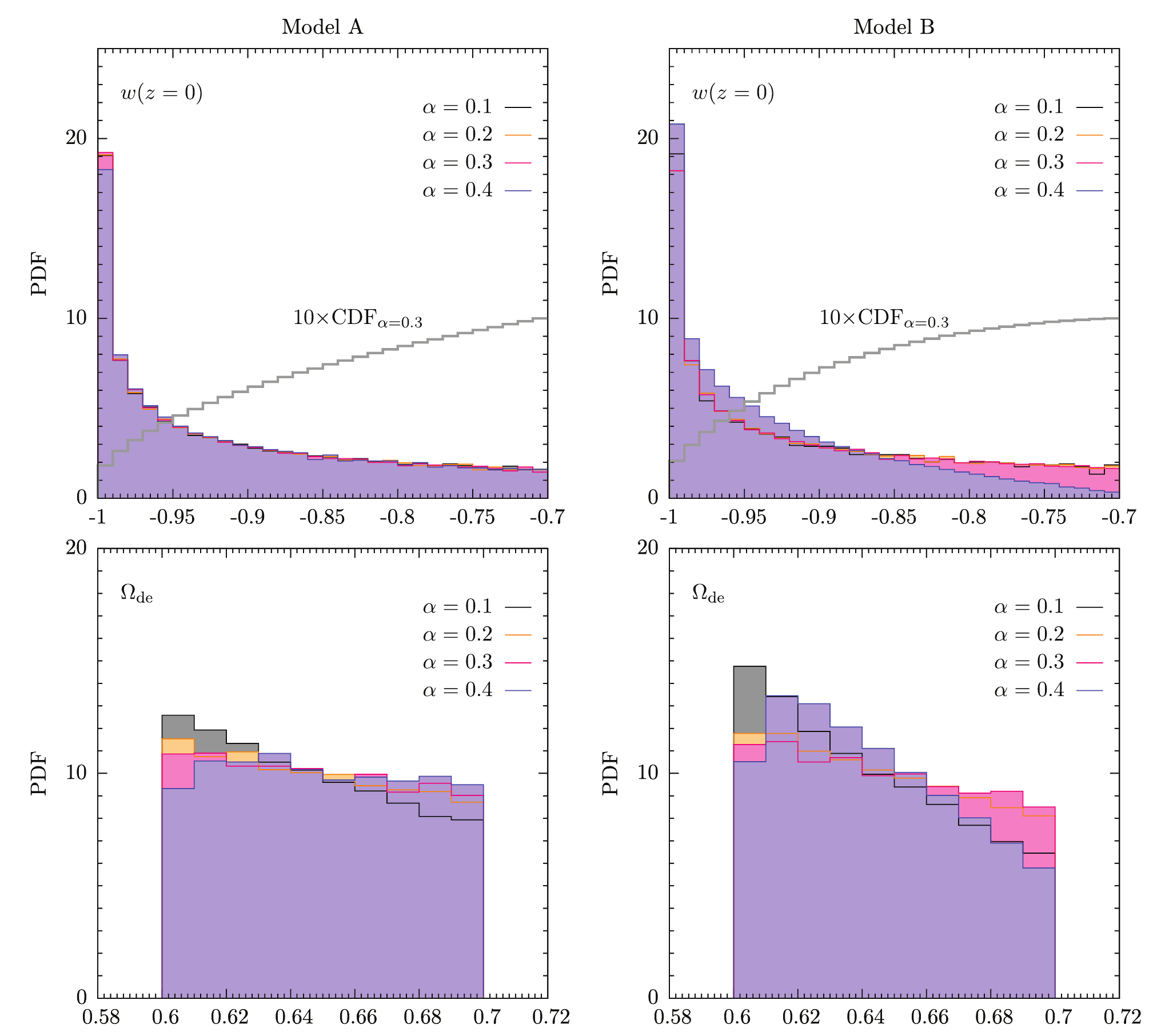}
\caption{The probability distribution functions (PDFs) for the dark energy equation-of-state parameter $w$ and density parameter $\Omega_{\rm de}$
 that arise in a family of
  models uniformly distributed in axion mass and initial misalignment
  angle which qualify as quintessence candidates by the criteria of 
  Sec.~\ref{sec:conditions} [\textit{i.e.}~$0.6\leq\Omega_{\rm de}\leq0.7$ and
  $w(0)\leq-0.7$]. Results are shown for four different values of
  $\alpha$. The gray lines show the cumulative probability distribution ($\times 10$) for $\alpha =0.3$ that
quantifies the probability of having $w$ near $-1$ among the dark
energy candidates; for example, $p(w\leq-0.95) \sim 40\%$.  The string axiverse scenario becomes less attractive in the sense that the scenario does not uniquely predict that $w \neq -1$.}
\label{fig:pdf}
\end{figure*}
\begin{figure*}[h!tb]
\centering
\includegraphics[width=6.1 in]{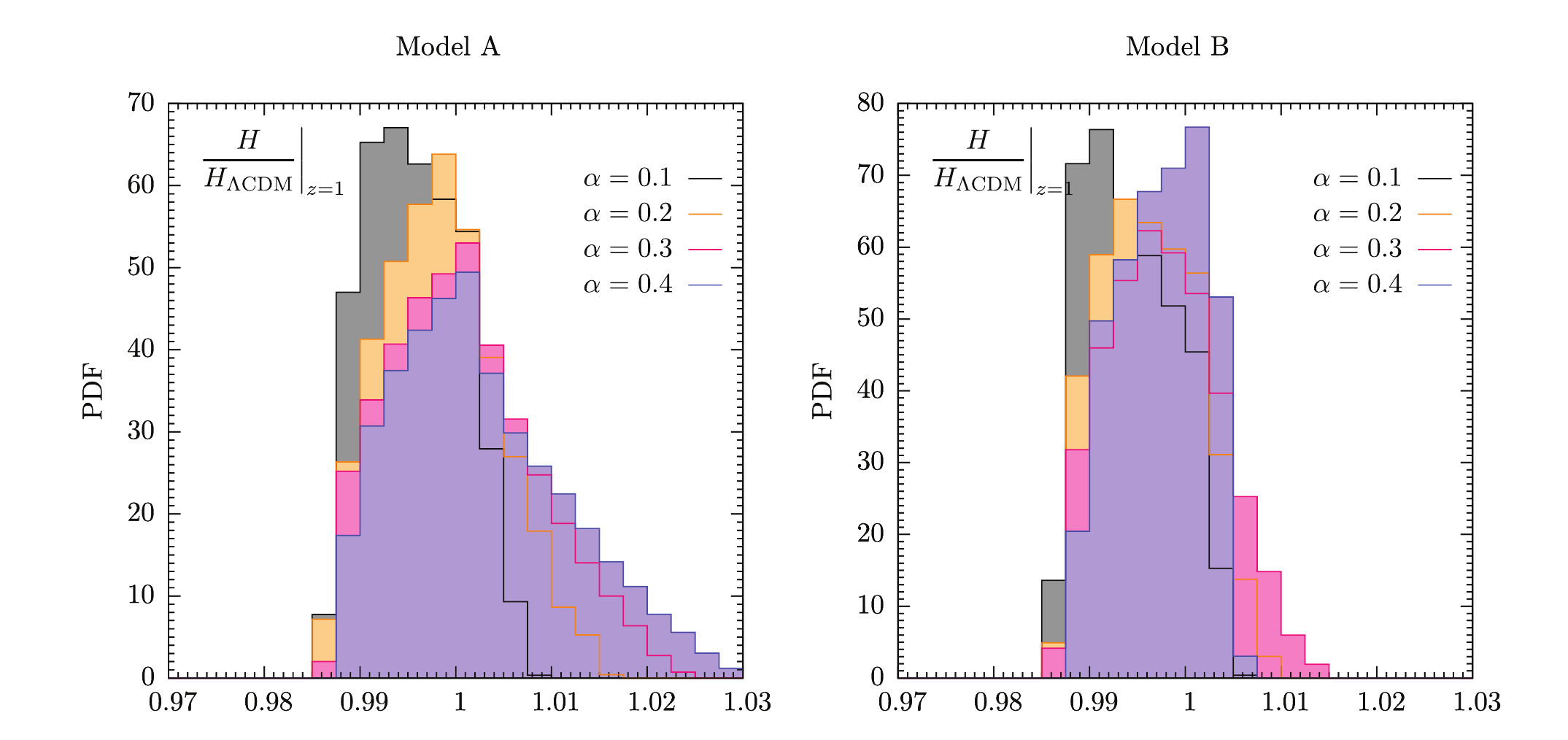}
\caption{The PDF for the Hubble parameter at redshift $z=1$ that arises in a family of
  models uniformly distributed in axion mass and initial misalignment
  angle which qualify as quintessence candidates by the criteria of 
  Sec.~\ref{sec:conditions} [\textit{i.e.}~$0.6\leq\Omega_{\rm de}\leq0.7$ and
  $w(0)\leq-0.7$]. Results are shown for four different values of
  $\alpha$.}
\label{fig:pdf_hubble}
\end{figure*}

\begin{figure*}[h!tb]
\centering
\includegraphics[width=6.1 in]{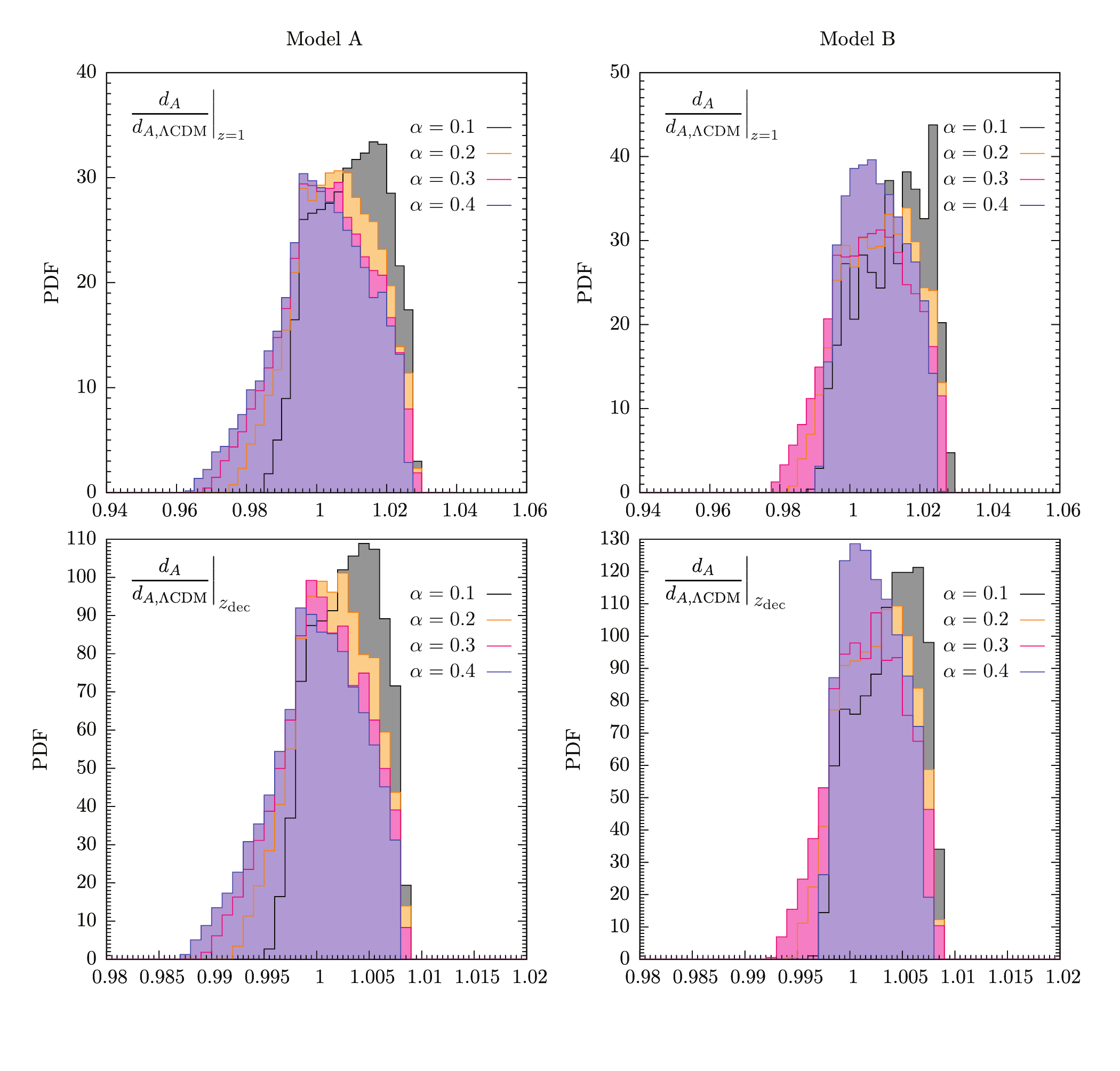}
\caption{The PDF for the angular-diameter distance $d_{A}$ to redshift $z=1$ that arises in a family of
  models uniformly distributed in axion mass and initial misalignment
  angle which qualify as quintessence candidates by the criteria of 
  Sec.~\ref{sec:conditions} [\textit{i.e.}~$0.6\leq\Omega_{\rm de}\leq0.7$ and
  $w(0)\leq-0.7$]. Results are shown for four different values of
  $\alpha$.}
\label{fig:pdf_dang}
\end{figure*}

In the top row of Fig.~\ref{fig:pdf} we show the PDFs for $w$ at
redshift $z=0$ that arise for both models A and B for several
different values of $\alpha$.  The results indicate that the
distribution in $w$ covers all values relatively evenly, but with an
increase as $w\to -1$ which is due to an integrable square-root
singularity at $w\to -1$. The gray lines show the cumulative
probability distribution ($\times 10$) for $\alpha =0.3$ that
quantifies the probability of having $w$ near $-1$ among the dark
energy candidates; for example, $p(w\leq-0.95) \sim 40\%$.  We thus conclude that if the empirically allowed window for $w$ continues to converge to an ever smaller range around $w = -1$, the string axiverse scenario becomes less attractive in the sense that the scenario does not uniquely predict that $w \neq -1$. 

The second row of Fig.~\ref{fig:pdf} shows the PDFs
for~$\Omega_{\rm de}$. They are flat, and do not favor a particular
value of $\Omega_{\rm de}$. Some qualitative understanding might be
gained as follows: For a given $\alpha$, all dark energy candidate
solutions have approximately the same mass $m$ that is small enough
that little evolution of the field has yet occurred, but big enough to
yield $0.6\leq \Omega_{\rm de}\leq0.7$. For example, in Model~A and in the
limit of neglecting the field evolution altogether, the energy density
is proportional to
$\rho_{\phi} \propto (2-\delta_i^2/2) m^2 \alpha^2$.  For constant
$m $ and $\alpha$, $\rho_{\phi}$ is hence distributed as $1/\sqrt{\rho_{\phi}}$
(the PDF of $\delta_i^2$) and a relatively flat distribution is not surprising. If $\delta$ and $m$ are both drawn from a uniform distribution, $\rho_{\phi}$ still has a relatively flat distribution ($\propto 1/\sqrt{\rho_{\phi}}$ away from the edges of the prior on $\delta$ and $m$). The same argument holds roughly for Model~B, although the field evolution is more dramatic for this scenario.

Ratios of the Hubble parameter $H(z=1)$ to $\Lambda$CDM values are shown in Fig.~\ref{fig:pdf_hubble}, while the analogous plots for $d_{A}(z=1)$ are shown in the upper row of Fig.~\ref{fig:pdf_dang}, along with the equivalent plots for $d_{A}(z_{\rm dec})$ in the lowest row. Common to all
PDFs is that they are clustered around the fiducial \LambdaCDM\ case (by construction)
and deviate from it by no more than $2\%$. Such deviations fall within the reach of completed surveys like BOSS \cite{Anderson:2013zyy} and the projected sensitivities of 
galaxy redshift surveys by WFIRST \cite{2015arXiv150303757S}, SKA \cite{Maartens:2015mra, Raccanelli:2015qqa}, and SPHEREx \cite{Dore:2014cca}. In the
next section we explore this sensitivity and derive constraints from
existing observations.

\section{Constraints from the baryon-photon sound horizon}
\label{sec:currentconstraints}

In the analysis above, we roughly (and conservatively) identified the dark energy parameter space as $-1\leq w\leq-0.7$ and
$0.6 \leq \Omega_{\rm de} \leq 0.7$, and derived predictions for
the distribution of prospective observables in that region. 
The actual allowed region of parameter space, however, can be constrained more
precisely, using quantities [like $d_A(z)$ and $H(z)$] which are more straightforwardly related to observables. When this is done, the allowed region in $(w,\Omega_{\rm de})$ parameter space is not rectangular, as we have approximated so far. 

In this section, we impose constraints to the
model and find the implications for the $\{\delta, m,\alpha\}$
parameter space using precise cosmological data. We treat the $\Lambda$CDM cosmology with Planck fiducial values as our null hypothesis \cite{Ade:2015xua} or fiducial cosmology. In particular, the cosmic sound horizon at decoupling (of the baryon-photon plasma) sets the scale of the BAO feature in the galaxy power spectrum, as well as the angular scale of the CMB acoustic peaks. Below, we use both quantities to test the possibility of axiverse-inspired quintessence models. 

Constraints to the Model A case were also obtained in Ref. \cite{Smer-Barreto:2015pla}. In that work the effect of quintessence perturbations was not considered, more restrictive priors on $\Omega_{\rm de}$ were used, and the degree of fine tuning in $\log({\delta})$ is not explicitly stated. Our work also goes beyond that work in exploring the power of future experiments.

Given a parameter selection, the model makes specific
predictions for the time evolution of $w$ that can be constrained.
If there is no energy transfer between the quintessence field and
another component (as assumed in this work), energy conservation
$d(\rho_{\phi}R^3) = - p_{\phi}dR^3$ implies,
\begin{align}
  \rho_{\phi}(z)/\rho_{\phi,0}  =  \exp \left[  3\int_0^z dz' \frac{1+w(z')}{1+z'}\right] . 
\end{align}
In the simplest case, $w$=const, the right-hand side reduces to
$(1+z)^{3(1+w)}$ and $w=-1$ corresponds to a cosmological
constant. We could impose constraints to our axiverse-inspired quintessence models by determining which $\delta$, $m$, and $\alpha$ values are consistent with constraints to $w$.
This approach, however, neglects the (model-dependent) dynamics of $w$ and would lead to biases in the results, as well as an underestimate of the error bars.
One could also consider a simple two-parameter model, with $w(z) = w_0 + w_a z /(1+z)$.
Here $w_a = -2 dw/d\ln R|_{z=1}$ is a parameter that encodes the time dependence of the quintessence field ~\cite{Linder:2002et}, but this parameterization is not accurate for dark energy models in general or axion models in particular. 

Instead, we compute $d_A(z)$ and $H(z)$ (which are more directly related to survey observables) at various redshifts for a grid of models in the $\left\{\delta,m,\alpha\right\}$ parameter space and compare with survey data. With the mapping of vast volumes of our universe, measurements of galaxy clustering and the BAO feature embedded in them have proved to be
a powerful tool for constraining the cosmic expansion history and various cosmological parameters (see e.g. Refs.~\cite{Peacock:2001uk, Percival:2004fs, Guzzo:2008ac, Blake:2011en, Samushia:2011cs, Reid:2012sw, Anderson:2013zyy, Samushia:2013yga, Raccanelli:2012gt, Gil-Marin:2015nqa}). The characteristic length scale that is imprinted by sound waves in the early Universe
can be computed as the comoving sound horizon
$r_d=\int_{z_d}^{\infty}c_s(z)/H(z) dz \simeq 150\,\Mpc $ at
the drag epoch with redshift $z_d$, not to be confused with the sound horizon $r_{\rm dec}$ at decoupling. By measuring the sound horizon in
the galaxy two-point correlation function, one can then infer $d_A(z)$ and $H(z)$ at low
redshift~$z$. Roughly speaking (for details, see the detailed discussion of Refs. \cite{Samushia:2011cs,Samushia:2011cs,Anderson:2013zyy} and references therein), galaxy separations along the line of sight depend on $H(z)r_d$
(corresponding to differences in redshift) and separations transverse
to the line of sight depend on $d_A(z)/r_d$ (corresponding to
differences in angular position).

We also compute the sound horizon $r_{\rm dec}=\int_{z_{\rm dec}}^{\infty}c_s(z)/H(z) dz \simeq 145\,\Mpc $ at the redshift of decoupling, $z_{\rm dec}\simeq 1090$. This quantity can be related to the angular sound horizon at decoupling \cite{Ade:2015xua}, \begin{align}
  \theta_{*} = \frac{1}{1+z_{\rm dec}} \frac{ r_{\rm dec} }{ d_A(z_{\rm dec}) }.
\end{align} This quantity sets the angular scale of acoustic peaks in the CMB acoustic peaks, and is one of the most precisely measured numbers in cosmology. 

By altering the late-time cosmic expansion history, axion dark energy will also change $d_{A}(z_{\rm dec})$, the angular-diameter distance to the surface of last scattering. This will alter the angular scale of the CMB acoustic peaks, yielding an additional CMB constraint to axion dark energy. For some model parameter values, the physical sound horizon $r_{\rm dec}$ at decoupling changes (very modestly), altering the dependence of the angular acoustic scale on axion model parameters. As we show in Appendix \ref{sec:cmb}, computing the effect of axiverse-inspired quintessence models on $\theta_{*}$ is a very accurate (and much faster) proxy for a full computation of CMB power spectra, and can thus be efficiently used to infer CMB constraints to such models.

\begin{figure*}[tb]
\includegraphics[width=\textwidth]{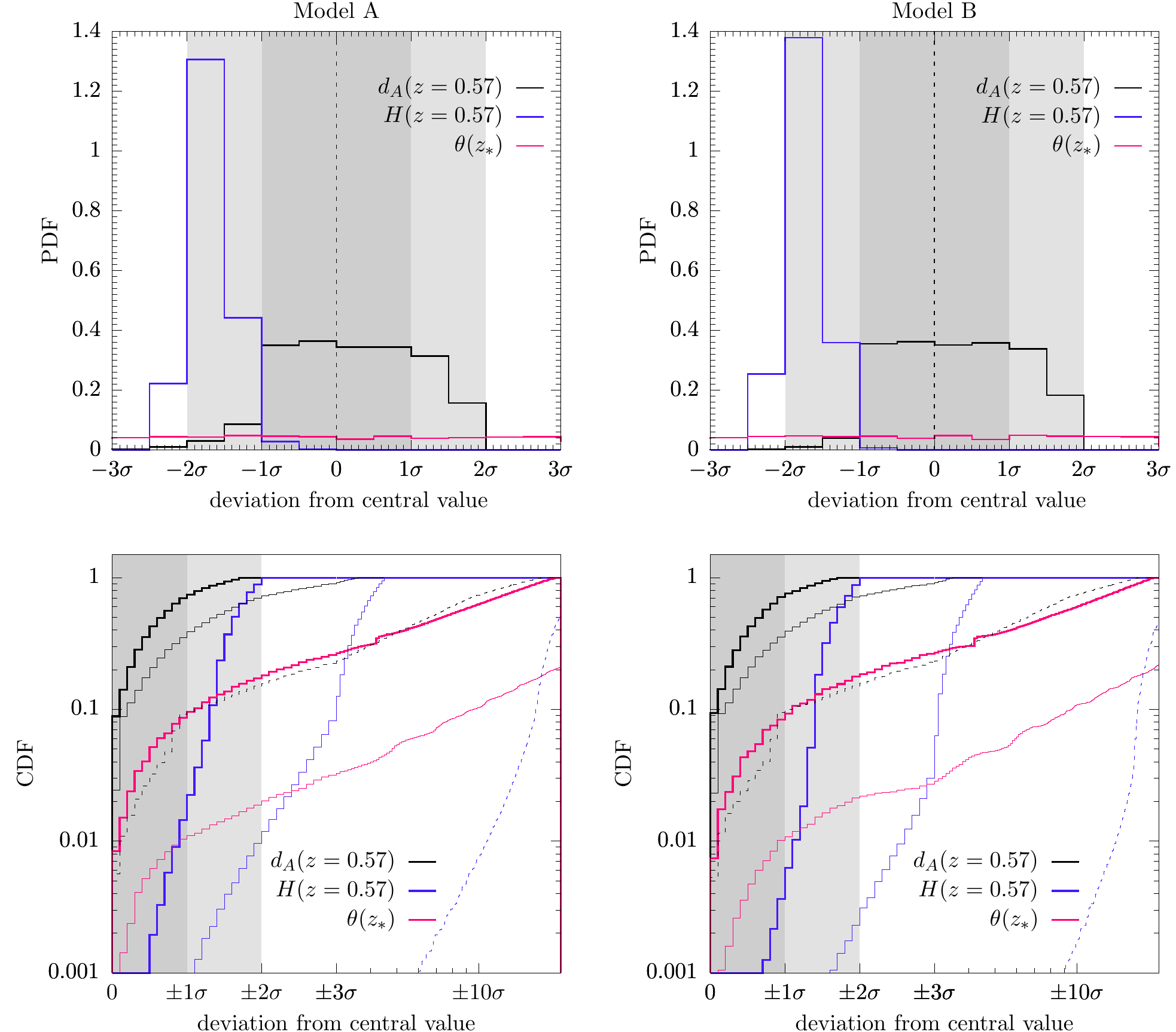}
\caption{Distributions of the observables $d_{A}(z=0.57)$ (solid black curves), $H(z=0.57)$ (solid blue curves), and $\theta_{*}$ (solid pink curves) in axiverse-inspired quintessence models, plotted as functions of the number of standard deviations from the central values stated in Eqs.~(\ref{eq:bao_constraint})-(\ref{eq:cmb_constraint}). The left column shows results for Model A, while the right column shows results for Model B. The top panel shows the PDF and only shows existing data (anisotropic BAO measurements from BOSS and CMB data from Planck). The bottom row shows the cumulative distribution function (CDF), now both comparing models with current data as in the first row, but also with future experiments, assuming a $\Lambda$CDM fiducial model. Errors are smaller for future experiments, leading to an outward dilation of the CDF (more models are ruled out with better data). In the second row, the light black curve shows predicted results for $d_{A}(z=0.57)$ with the SPHEREx experiment. The dotted black curve shows predicted results for $d_{A}(z=0.57)$ with the Square Kilometre Array (SKA). The light blue curve shows predicted results for $H(z=0.57)$ with SPHEREx. The dotted blue curve shows predicted results for $H(z=0.57)$ with the SKA. The thin pink curve shows predicted results for a cosmic-variance limited CMB polarization experiment with maximum multipole $\ell_{\rm max}=2200$, similar to the CMB-S4 concept.}
\label{fig:hist}
\end{figure*}

We now quantify the discriminating power of the various cosmological observables for
the axiverse-inspired models that satisfy the basic requirements
$w\leq -0.7$ and $0.6\leq \Omega_{\rm de} \leq 0.7$. We obtain constraints from current data, and then assess the sensitivity of future experiments.
Figure~\ref{fig:hist} shows histograms [probability distribution functions (PDFs) and cumulative distribution functions (CDFs)] of the predictions of the
models and how they fare in comparison to BOSS observed values of
$d_A$ and $H$ at redshift $z=0.57$ and to the Planck 2015 inferred
angular scale of the sound horizon at decoupling.

For the central values and errors of the BAO
observables we use measurements of the angular diameter distance to and Hubble parameter at $z = 0.57$ from the measurement of the BAO peak in the correlation of galaxies from the Sloan Digital Sky Survey III Baryon Oscillation Spectroscopic Survey~\cite{Anderson:2013zyy},
\begin{align}
 d_A(z=0.57) & =  (1421 \pm 20)\,{\rm Mpc} \times(r_d/r_{d,\rm fid}) , \nonumber \\
 H(z=0.57) & =  (96.8\pm 3.4)\,{\frac{\rm  km/sec}{\rm Mpc}} \times (r_{d,\rm fid}/r_d) ,\label{eq:bao_constraint}\end{align} where $r_{d,\rm fid}=149.28~{\rm Mpc}$ \cite{Anderson:2013zyy} and $r_{d}=147.43~{\rm Mpc}$ for our assumed $\Lambda$CDM fiducial model. For other models in the axion parameter space $r_{\rm d}$ takes on different values and we compute those self-consistently. We include the covariance between these quantities \cite{Anderson:2013zyy} when determining if a given axion model is inside or outside the allowed region at the observable level. For Planck, we are conservative and use the value
 \begin{align}
 100 \theta_{*} & = 1.04105 \pm 0.00046  \label{eq:cmb_constraint},
\end{align} stated in Ref.~\cite{Ade:2015xua}, based on temperature data at all scales and low $\ell$ polarization data (as opposed to the full polarization data set).
For Model A, we swept the parameter space on a logarithmic grid satisfying $0.001\leq\delta_{i}\leq2$ and $0.1\leq\alpha\leq0.8$, with 81 $\delta_{i}$ values and $31$ $\alpha$ values. We then obtain central $\overline{m}(\delta, \alpha)$ value for the axion mass by demanding that $\Omega_{\phi}=1$ in the no-roll limit, which implies that $\overline{m}/H_{0}=1/\alpha$. We then define a scan range of $61$ $m$ values, linearly spaced with $m/H_{0}=(\overline{m}/H_{0})^{+0.6}_{-0.4}$. For Model B, the same density condition implies that $\overline{m}/H_{0}=1/(2\alpha)$, with $m/H_{0}=(\overline{m}/H_{0})\pm 0.3$. Here we use the range $10^{-6}\leq \delta_{i}\leq 1$, with the same number of grid points as for Model A, in all dimensions. We check that these values ensure a full sweep of the allowed parameter space.

We ran a full numerical integration of the equations of motion [Eqs.~(\ref{eqn:EOM}) and (\ref{eqn:hubble})] for all $153,171$ models in this grid and then compare the values of $d_{A}(z)$, $H(z)$ and $\theta_{*}$ with the BOSS and Planck confidence regions described above. We also ran this integration within a modified version of the \textsc{camb} code, described in Appendix \ref{sec:cmb}, in order to self-consistently compute the (small) change in $\theta_{*}$ in our model that results from modified early-time evolution of the (highly sub-dominant) dark-energy component, including small corrections to the value of $z_{\rm dec}$. We found that our results are insensitive to this correction. 

Constraints from current measurements and predictions of future sensitivity are shown in Fig. \ref{fig:hist}. For the duration of this (parameter-space volume) discussion, $d_{A}(z)$ are $H(z)$ are treated simply, as uncorrelated independent observables. Correlations are considered more carefully below, when detailed parameter constraints to $\left\{\delta,m,\alpha\right\}$ are obtained. 

For Model A (B), at the $1\sigma$ confidence level, $70\%$ ($ 71\%$) of the models (in our grid) consistent with dark energy priors are allowed by the $d_{A}(z=0.57)$ measurement, while only $1.4\%$ ($0.4\%$) of the models are allowed by the $H(z=0.57)$ measurement. This is driven by the well known $1\sigma$ tension between BOSS results and Planck fiducial $\Lambda$CDM values \cite{Anderson:2013zyy}. 

For Model A (B), at the $2\sigma$ confidence level, $99\%$ (all) of the models consistent with the dark energy prior ($-0.7\geq w \geq-1$ and $0.6\leq \Omega_{\phi}\leq 0.7$) are allowed by the $d_{A}(z=0.57)$ measurement, while $89\%$ ($87\%$) of the models are consistent with the $H(z=0.57)$ measurement. Finally, CMB measurements are more discriminating. For Model A (B), at the $2\sigma$ confidence level, only $17\%$ ($18\%$) of the models are consistent with Planck constraints.

In addition to the high precision BAO measurements at $z=0.57$ from galaxy clustering, there are BOSS Lyman-$\alpha$ forest (LyaF) data that test the expansion history at higher redshift as well. The latter data show a $(2-2.5)\sigma$ tension with a CMB-supported standard cosmology, with LyaF results favoring a lower Hubble rate and smaller angular diameter distance at $z=2.34$ \cite{Aubourg:2014yra}. It is worthwhile to consider whether or not axiverse-inspired quintessence models can relieve that tension, while preserving agreement with BAO data at z = $0.57$. It turns out that they cannot, as we obtain no data points (for Model A or B) that agree with LyaF (Fig. 15 of Ref. \cite{Aubourg:2014yra}) constraints and measurements of $d_{A}(z=0.57)$ and $H(z=0.57)$. Of course this tension could be a statistical fluke, or the result of some unrecognized systematic error \cite{Aubourg:2014yra}.

Assuming a fiducial Planck $\Lambda$CDM cosmology, we also determine how future galaxy surveys could improve on those constraints, by performing a standard Fisher forecast~\cite{Tegmark:1996bz} of the errors on the observed galaxy power spectrum (at $z=0.57$) for the proposed SPHEREx satellite \cite{Dore:2014cca} and the Square Kilometre Array (SKA) \cite{Maartens:2015mra, Raccanelli:2015qqa} (at $z=1$). These results are obtained from a standard BAO forecast prescription \cite{Seo:2003pu} and are shown in the bottom panel of Fig. \ref{fig:hist}. 

For Model A and B, at the $2\sigma$ confidence level, we see that future $d_{A}(z=0.57)$ measurements from SKA will be comparably sensitive to current CMB constraints, potentially distinguishing all but $15\%$ of the parameter-space volume (allowed by our prior) from the $\Lambda$CDM model. The next major improvement in sensitivity will come from SPHEREx constraints to $H(z=0.57)$, which for Model A (B) will distinguish all but $1\%$ ($0.2\%$) of the parameter-space volume (allowed by our prior) from the $\Lambda$CDM model.

A future cosmic-variance limited CMB experiment (like CMB Stage-4 \cite{Abazajian:2013oma}) with Fisher-level sensitivity of $\Delta \theta_{*}/\theta_{*}\simeq 5\times 10^{-{5}}$ could observationally distinguish (from the $\Lambda$CDM model) a comparable parameter-space volume  (at $2\sigma$ confidence-level) to SPHEREx measurements of $H(z=0.57)$. The analogous measurement using the SKA could empirically distinguish (from the $\Lambda$CDM model) the whole parameter-space volume allowed by our prior, up to sampling uncertainties, which allow $1/10^{5}$ of the parameter-space volume to potentially evade constraints. Forthcoming cosmological observations could thus either validate axiverse-inspired quintessence models or show that they are extremely fine-tuned at the $1/10^{5}$ level (or worse) in terms of parameter-space volume.
%*** FIGURE ***
\begin{figure*}[tb]
\includegraphics[width=0.49\textwidth]{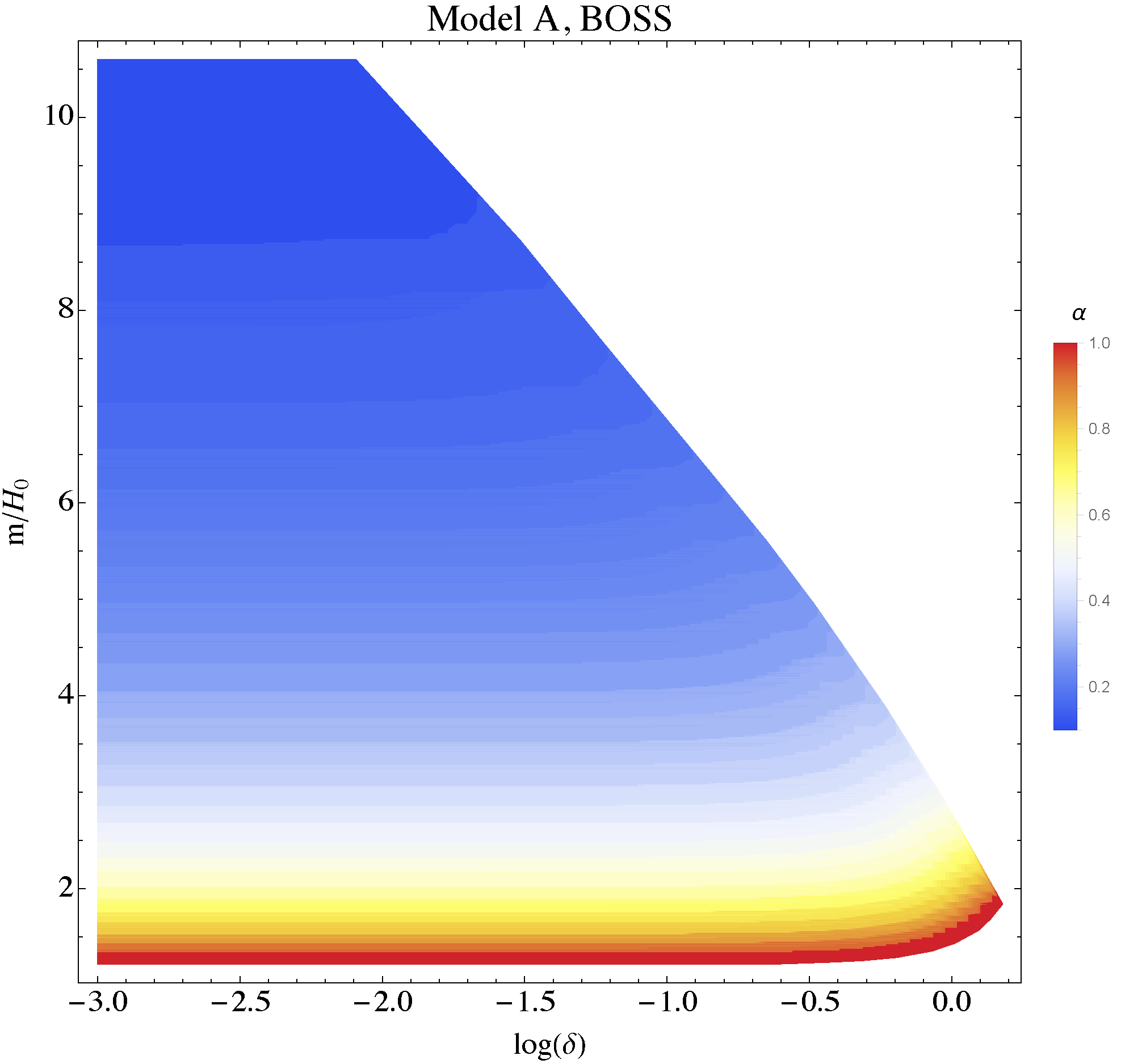}
\includegraphics[width=0.49\textwidth]{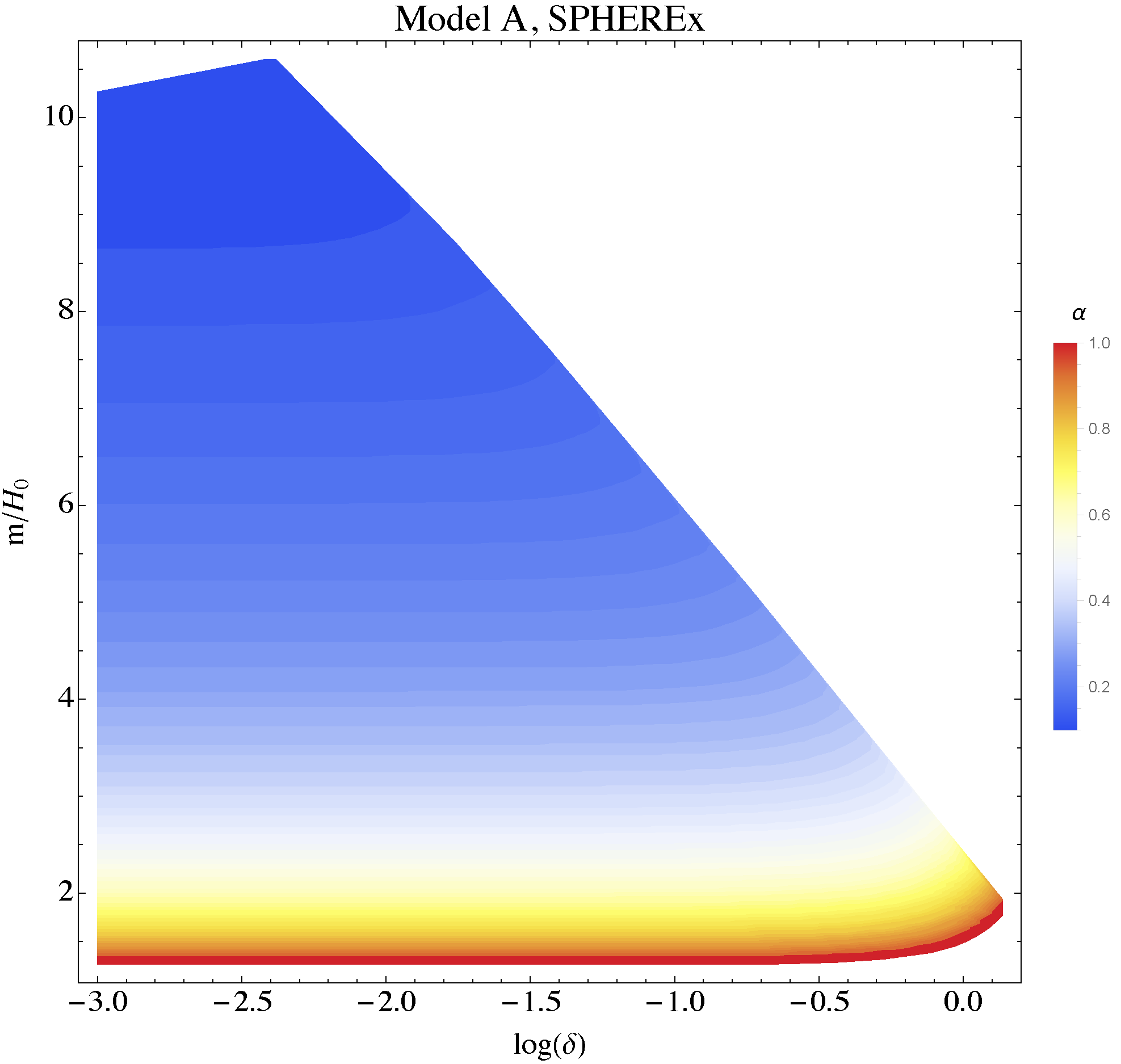}
\includegraphics[width=0.49\textwidth]{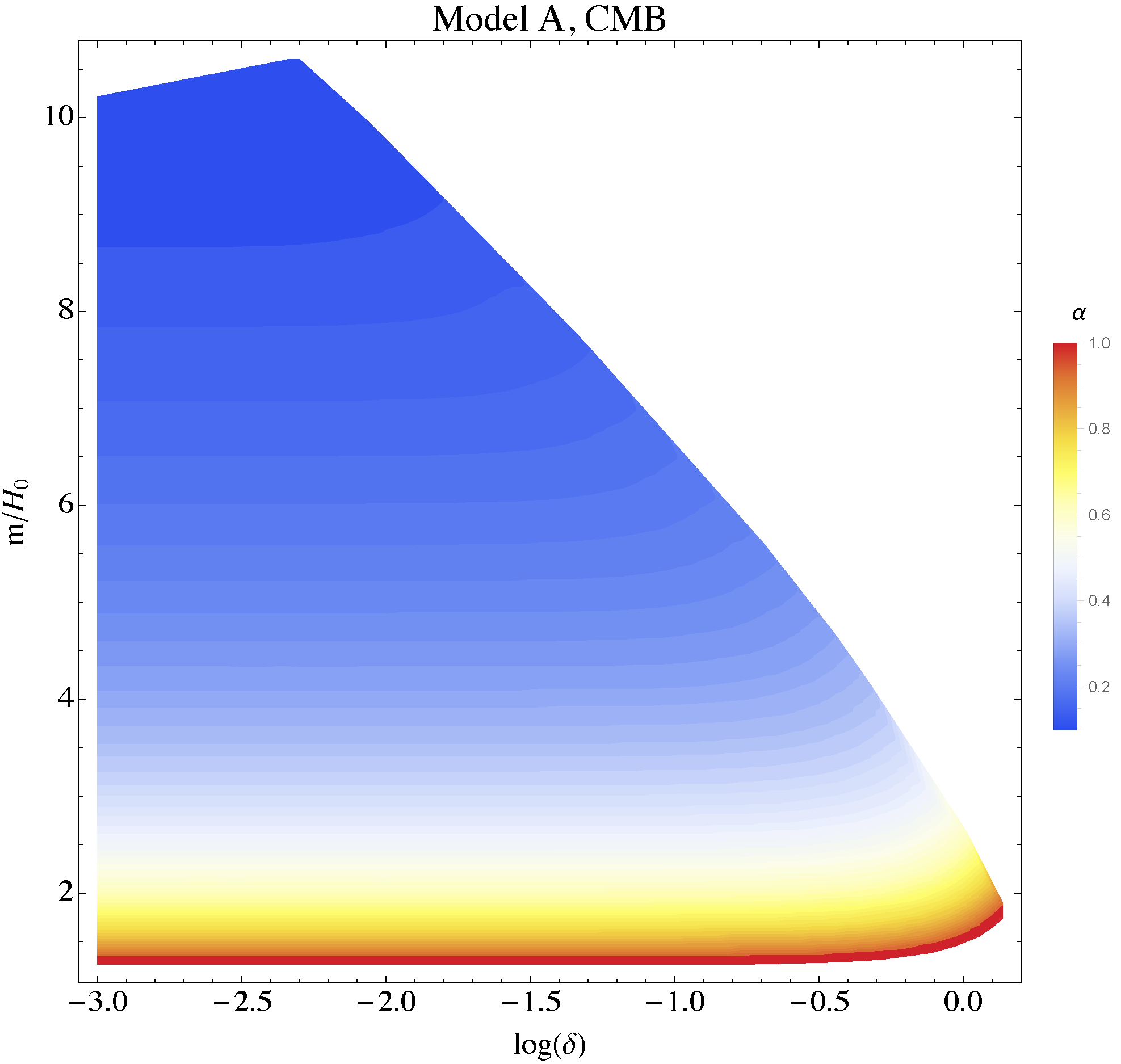}
\includegraphics[width=0.49\textwidth]{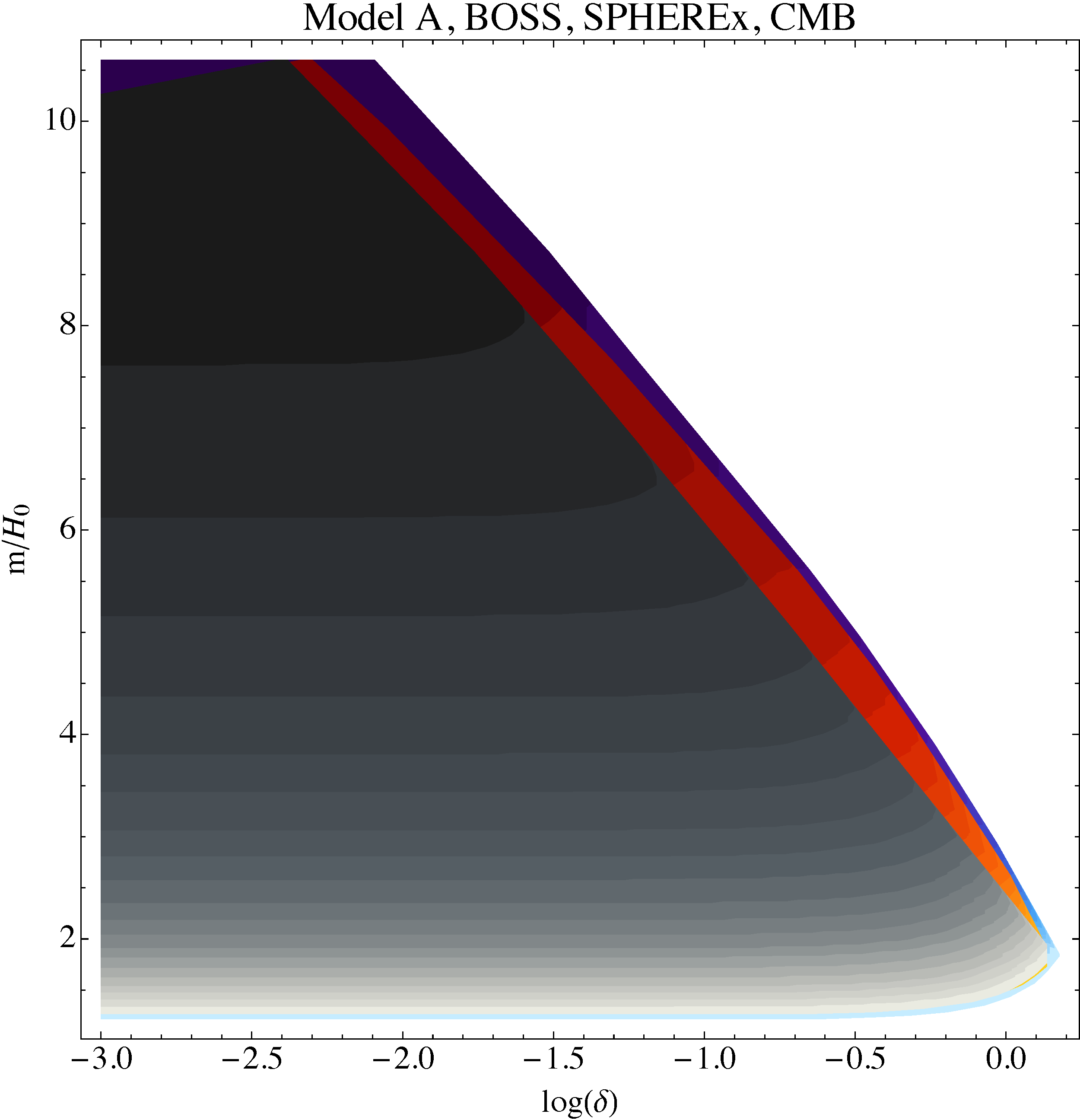}
\caption{Values of $\{\delta, m, \alpha\}$ for Model A allowed by different measurements or forecasts. Here and throughout this paper, $\log{(\delta)}$ refers to a base-$10$ logarithm.{\it Top left panel}: Constraints from the anisotropic BOSS BAO measurement of $\{d_A(z), H(z)\}$ at $z=0.57$ at 95\%~confidence level.
{\it Top right panel}: Same as top left panel, assuming the BOSS measurement holds correct, but with predicted error bars for SPHEREx.
{\it Bottom left panel}: Same as top panels but for measurements of $\theta_{*}$ from Planck.
{\it Bottom right panel}: Combination of the three previous panels. The blue contours show constraints from BOSS measurements. The red contours show Planck CMB constraints, while the black contours show forecasted constraints from SPHEREx.}
\label{fig:D_A-H_z057-CMB_bossfid}
\end{figure*}
%%
%
%%*** FIGURE ***
\begin{figure*}[tb]
\includegraphics[width=0.49\textwidth]{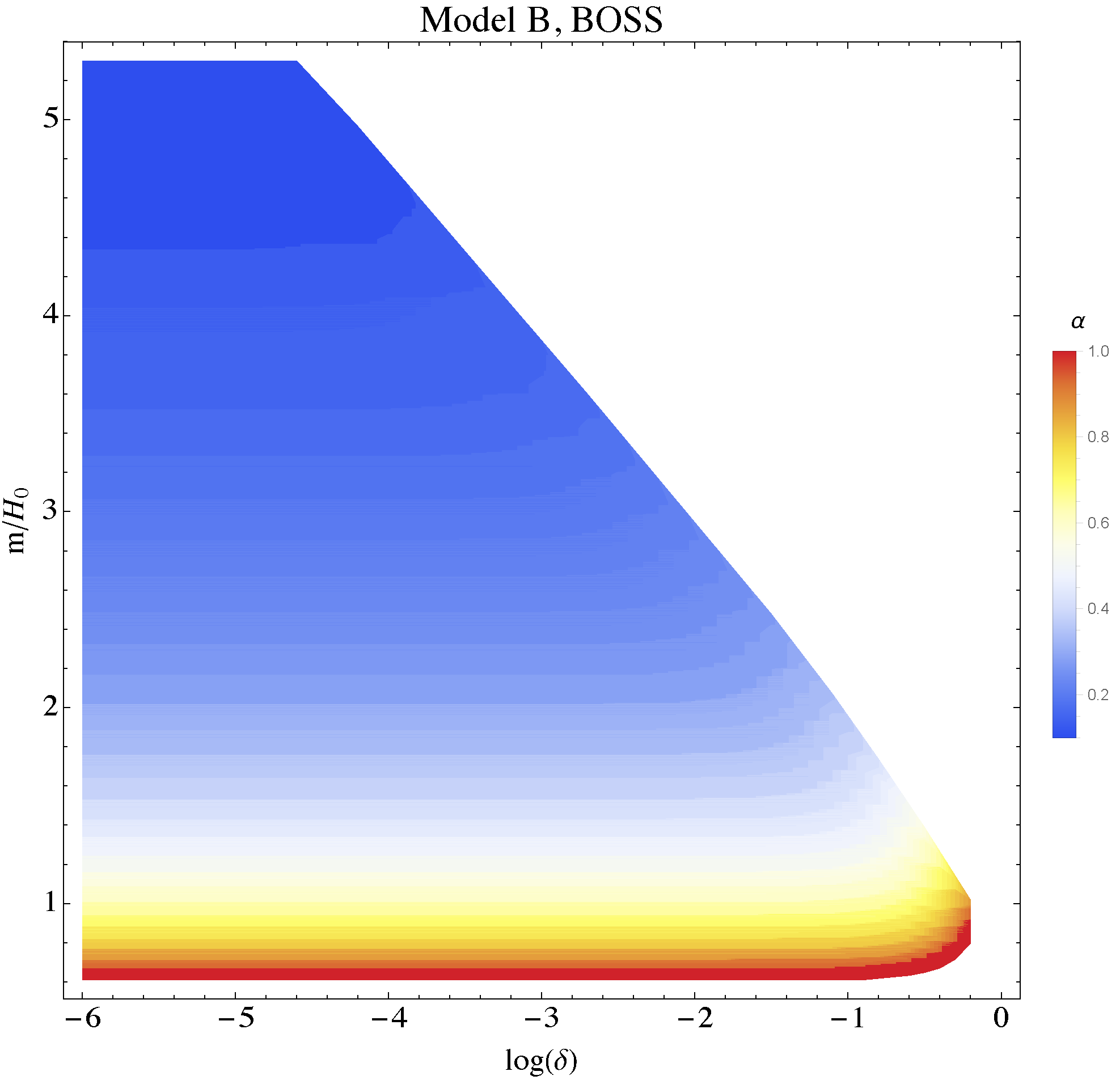}
\includegraphics[width=0.49\textwidth]{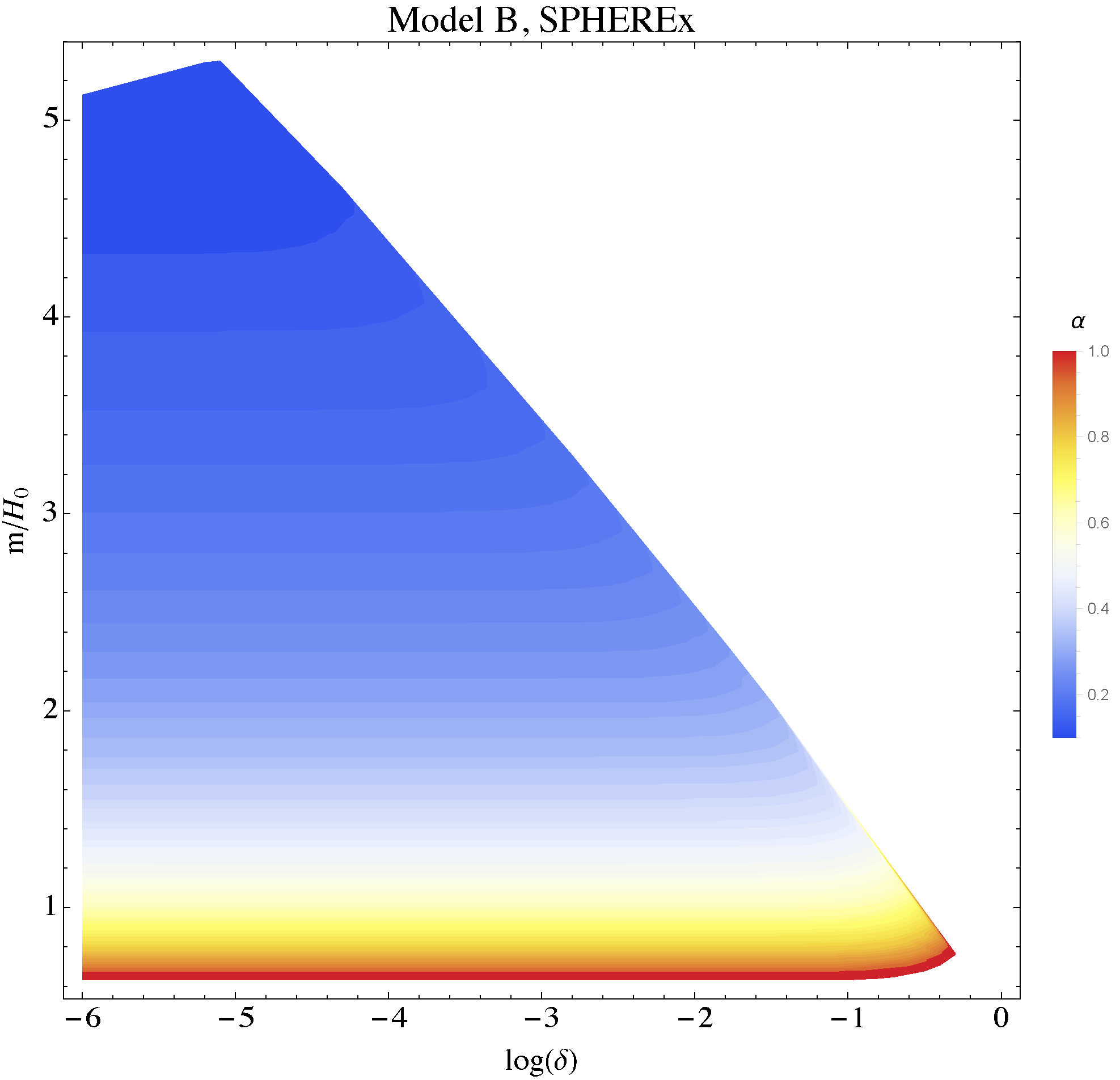}
\includegraphics[width=0.49\textwidth]{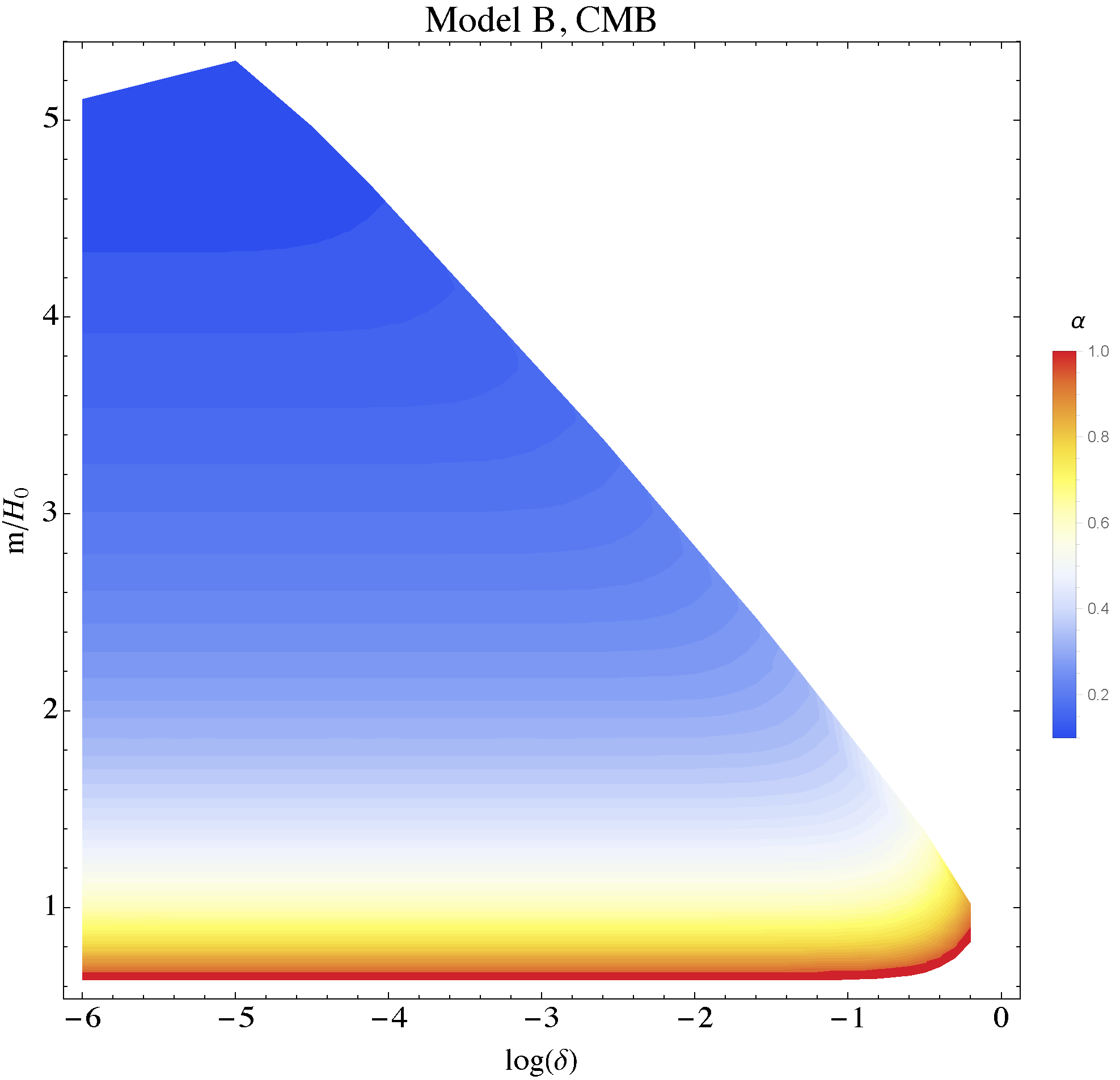}
\includegraphics[width=0.49\textwidth]{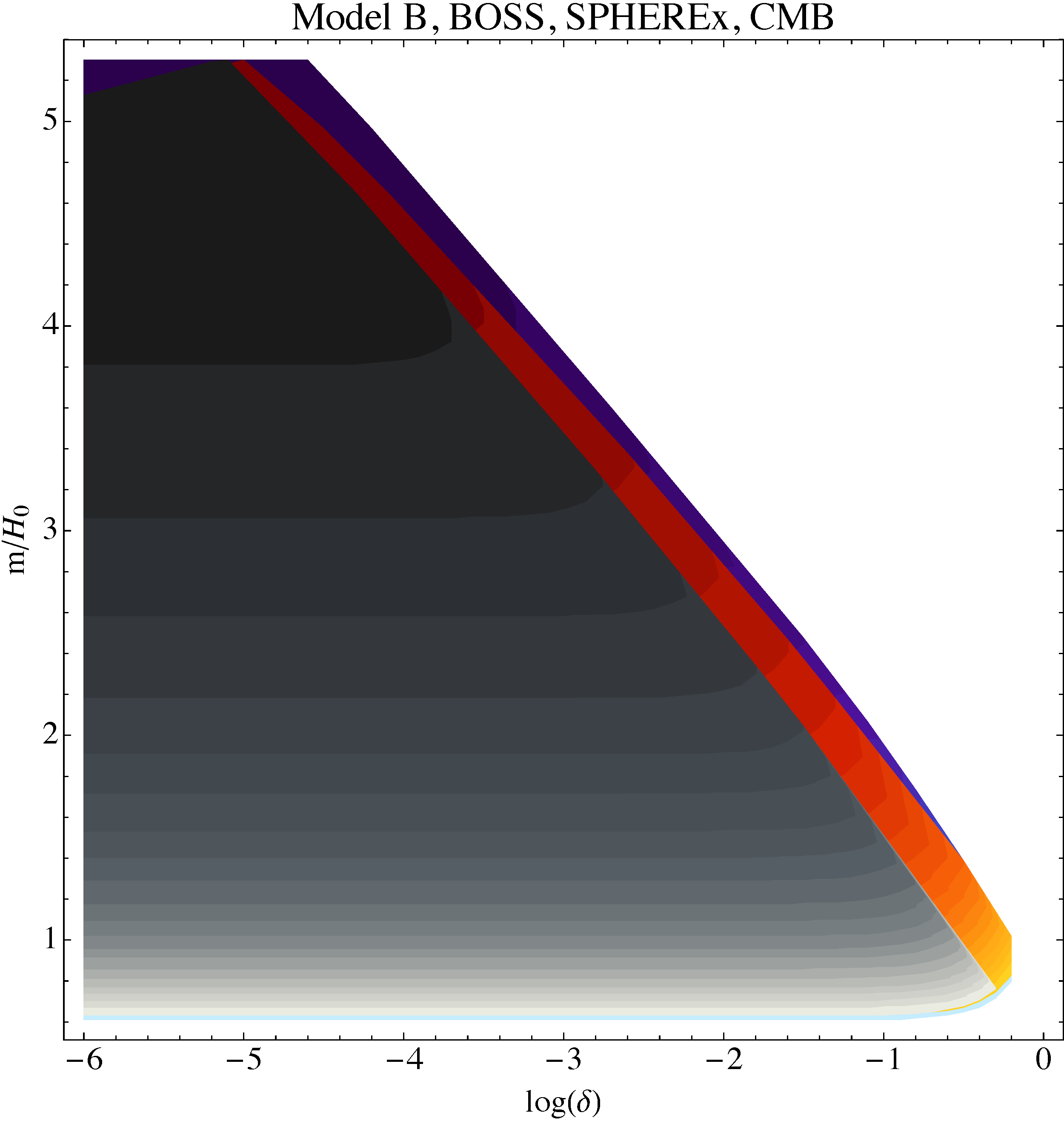}
\caption{Same color code/axes as in Fig.~\ref{fig:D_A-H_z057-CMB_bossfid} but for Model B.
}
\label{fig:D_A-H_z057-CMB_bossfid_B}
\end{figure*}

We now go beyond raw volume measurements to examine the detailed constraints in $\{\delta, m,\alpha\}$ parameter space. Here we show the allowed parameter space at $95\%$~confidence level. Results here are shown without the $0.6\leq\Omega_{\phi}\leq0.7$, $ w\leq -0.7$, `dark energy' prior, to see how much the data alone constrain the quintessence parameter space. The allowed region for $m$ sits well within the scan range described above, so even though this scan range amounts to a nontrivial (but linearly flat) prior, our conclusions do not depend on this prior. We then resort the constraints as a function $\alpha(m,\delta_{i})$ and use a third-order polynomial fit to densely interpolate between points to get (approximately) continuous physical allowed regions. 

We consider the same set of current/future experiments, with the addition of the WFIRST satellite \cite{2015arXiv150303757S} (considering a measurement of the BAO feature at $z=2$).
\begin{figure*}[tb]
\includegraphics[width=0.49\textwidth]{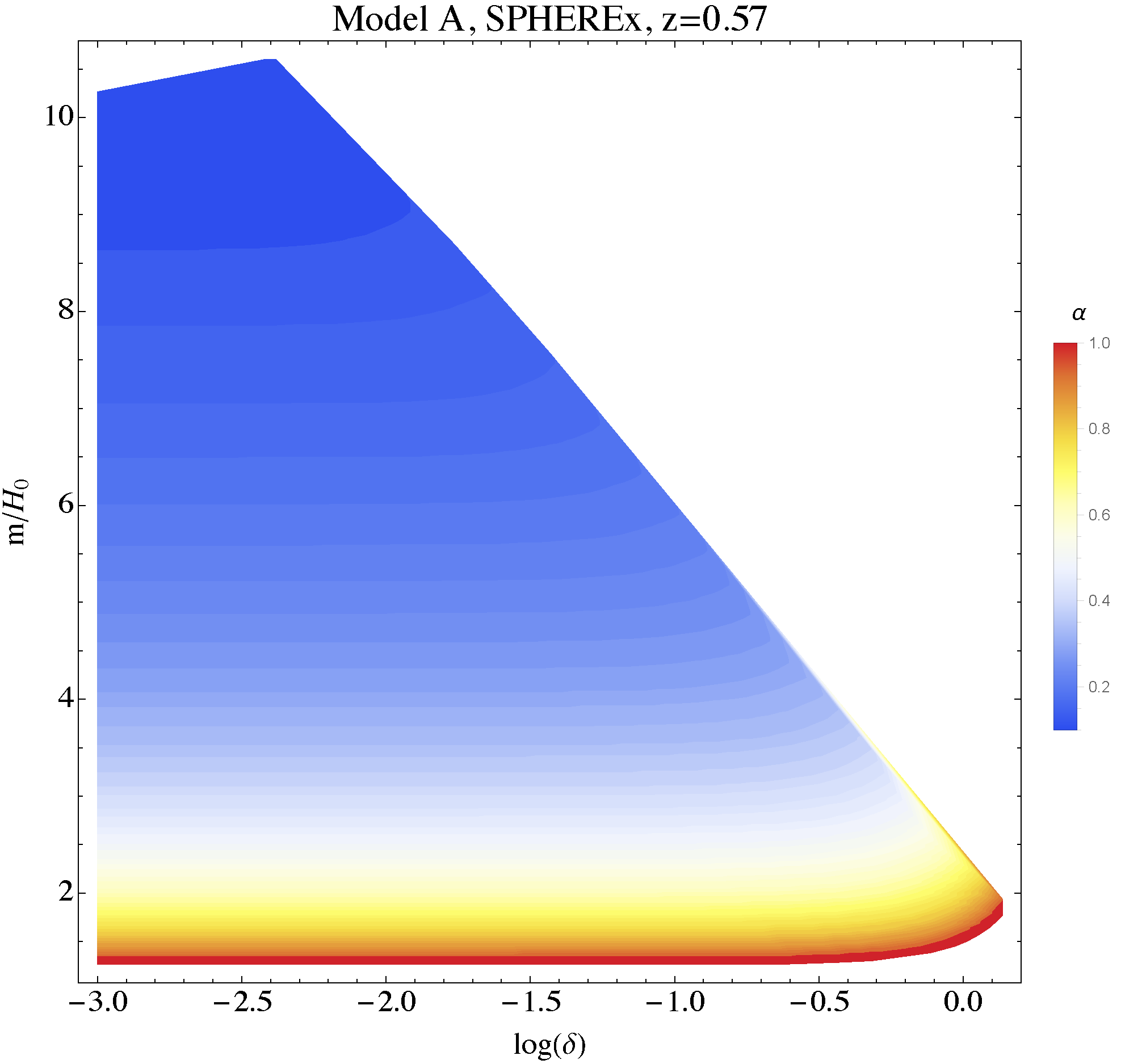}
\includegraphics[width=0.49\textwidth]{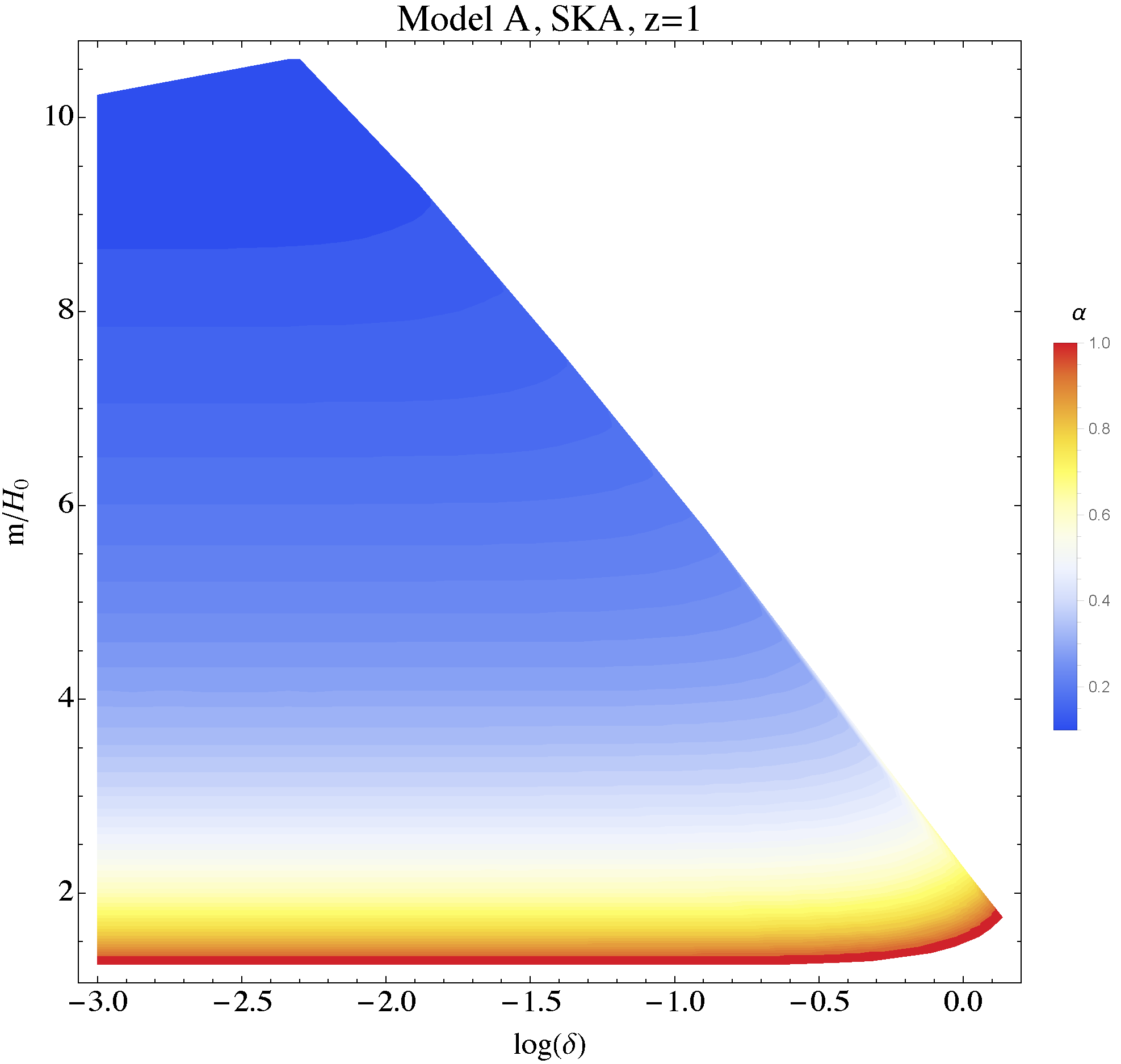}
\includegraphics[width=0.49\textwidth]{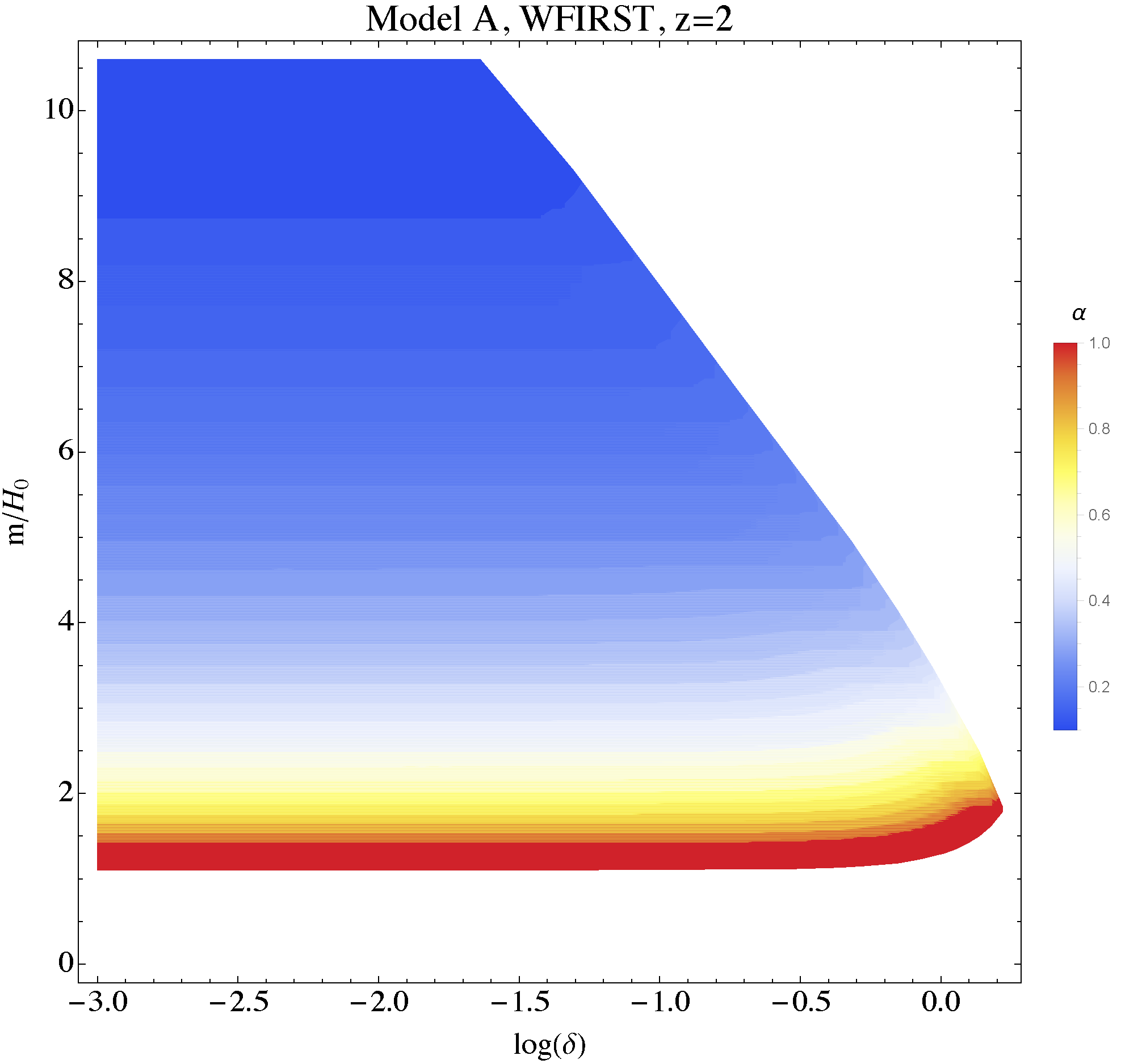}
\includegraphics[width=0.49\textwidth]{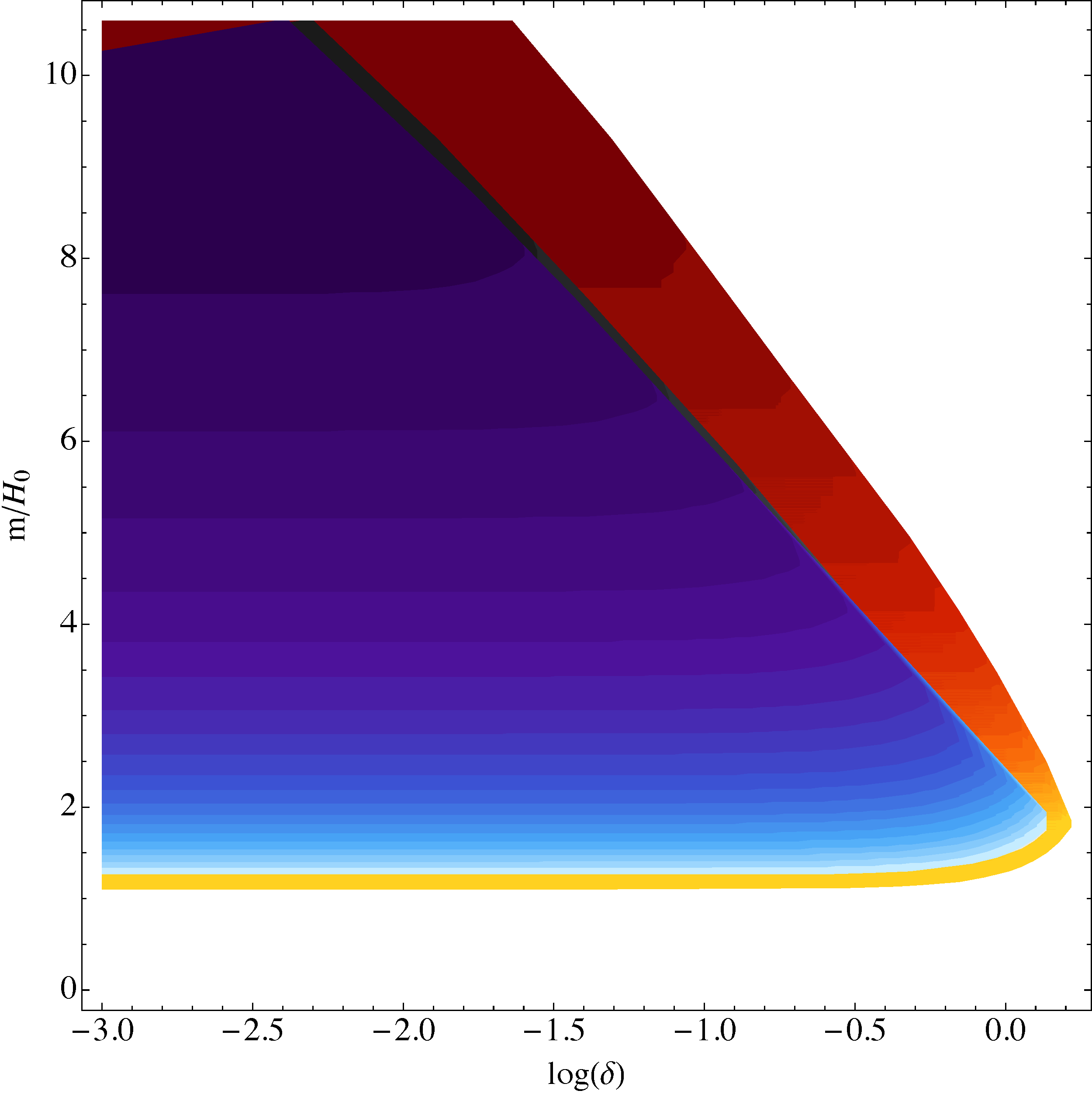}
\caption{Values of $\{\delta, m, \alpha\}$ for the Model A allowed by different forecasts, assuming the null hypothesis.e
Predicted constraints from future BAO measurements of $\{d_A(z), H(z)\}$.
{\it Top left panel}: Predicted constraints from SPHEREx measurements at $z=0.57$,
{\it Top right panel}: Predicted constraints from SKA measurements at $z=1$, 
{\it Bottom left panel}: Predicted constraints from WFIRST measurements at  $z=2$. 
{\it Bottom right panel}: Comparison of all constraints. In the top left, bottom left, and top right panels, colors indicate the value of $\alpha$, while in the comparison plot (bottom right panel), SPHEREx allowed regions are shaded in black, SKA in blue, and WFIRST in red.}
\label{fig:future}
\end{figure*}
\begin{figure*}[tb]
\includegraphics[width=0.49\textwidth]{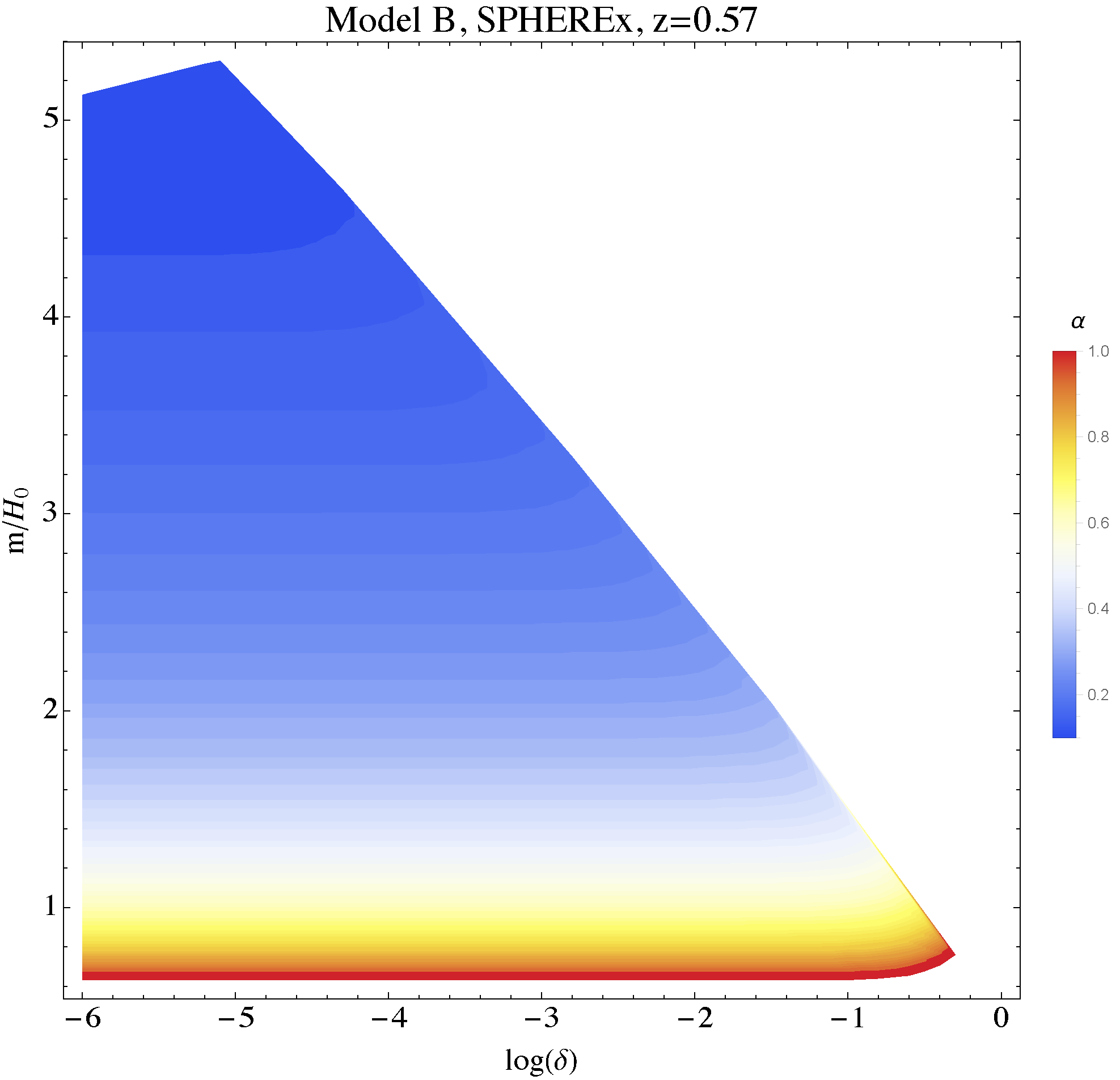}
\includegraphics[width=0.49\textwidth]{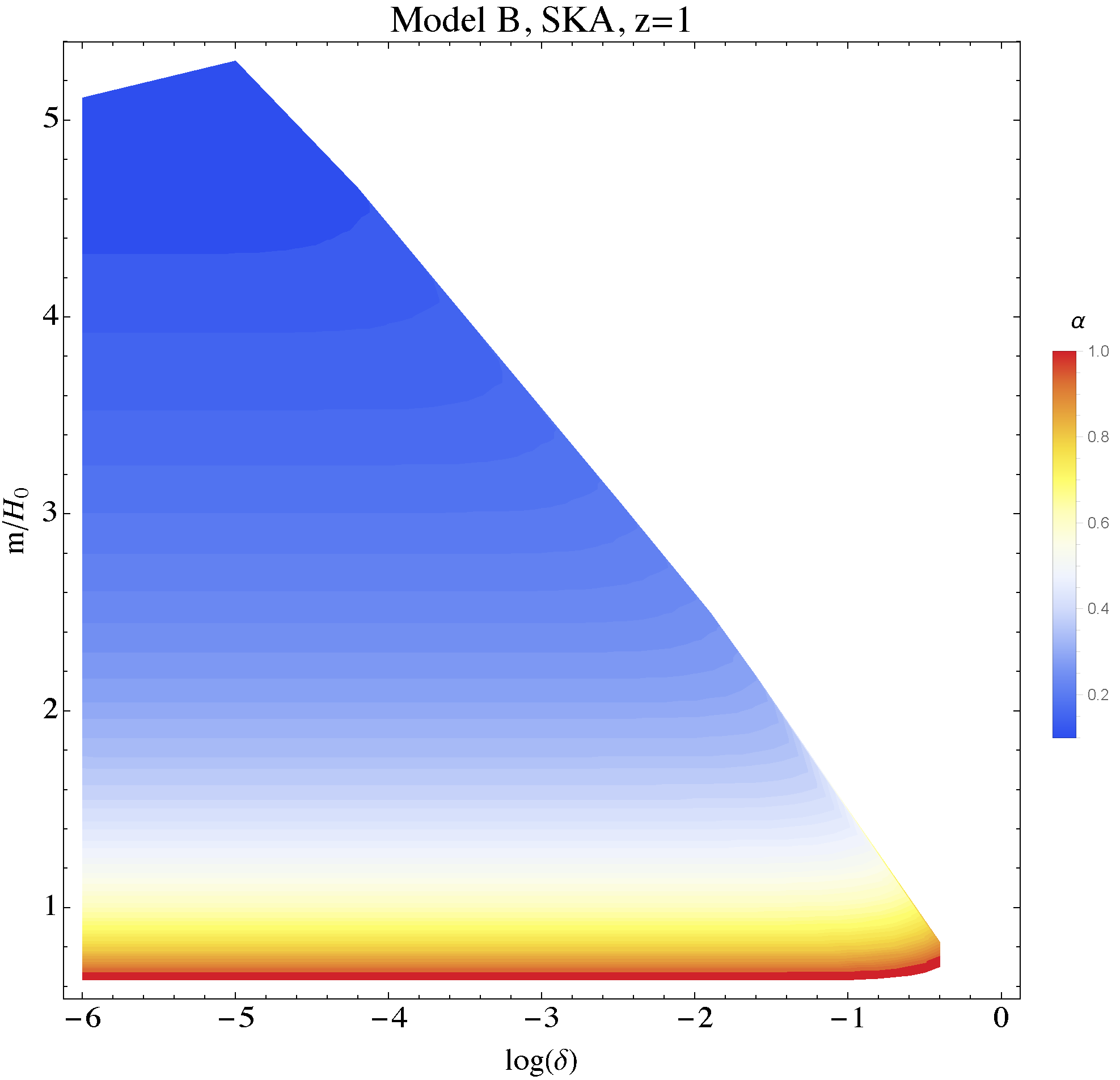}
\includegraphics[width=0.49\textwidth]{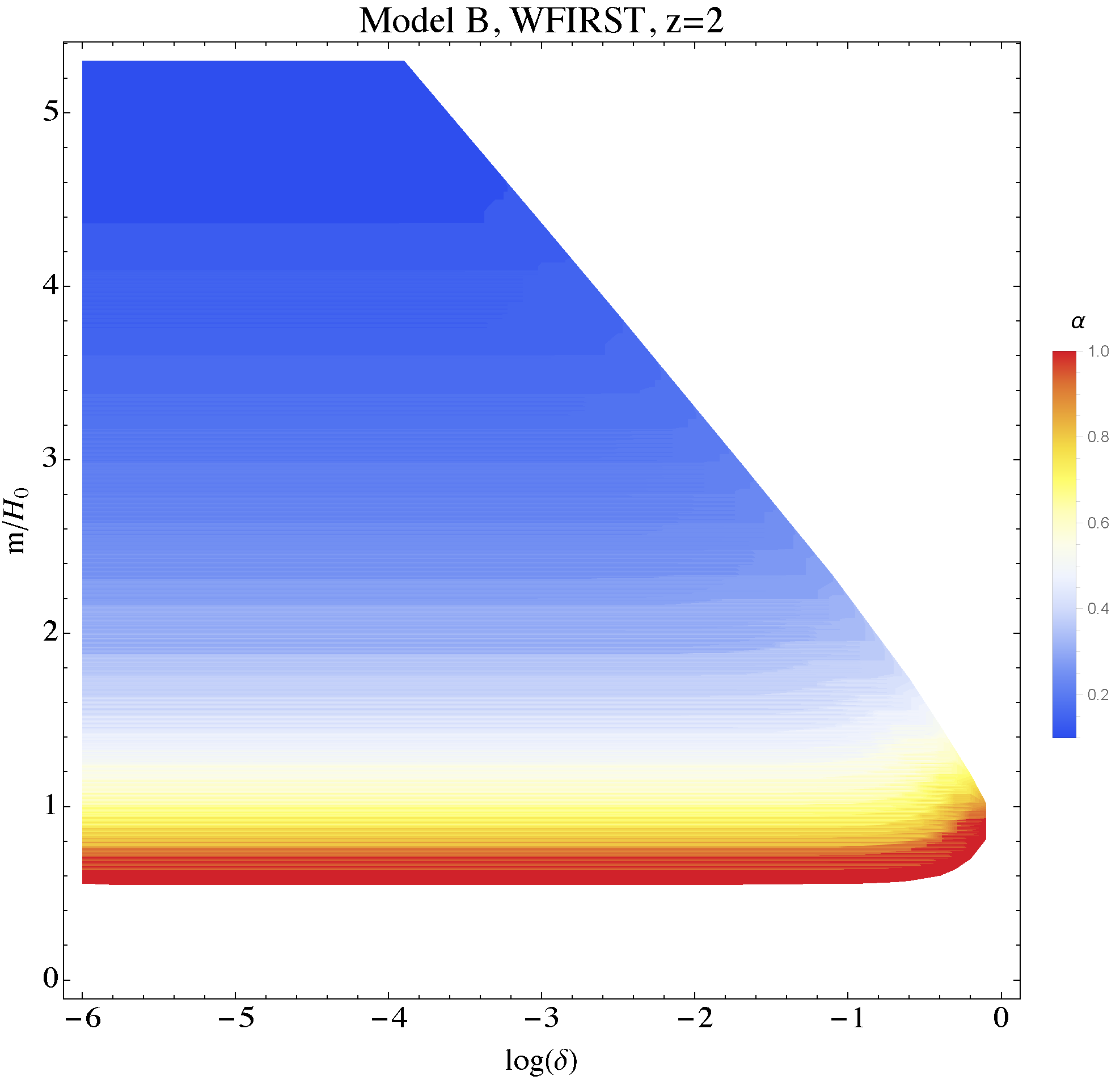}
\includegraphics[width=0.49\textwidth]{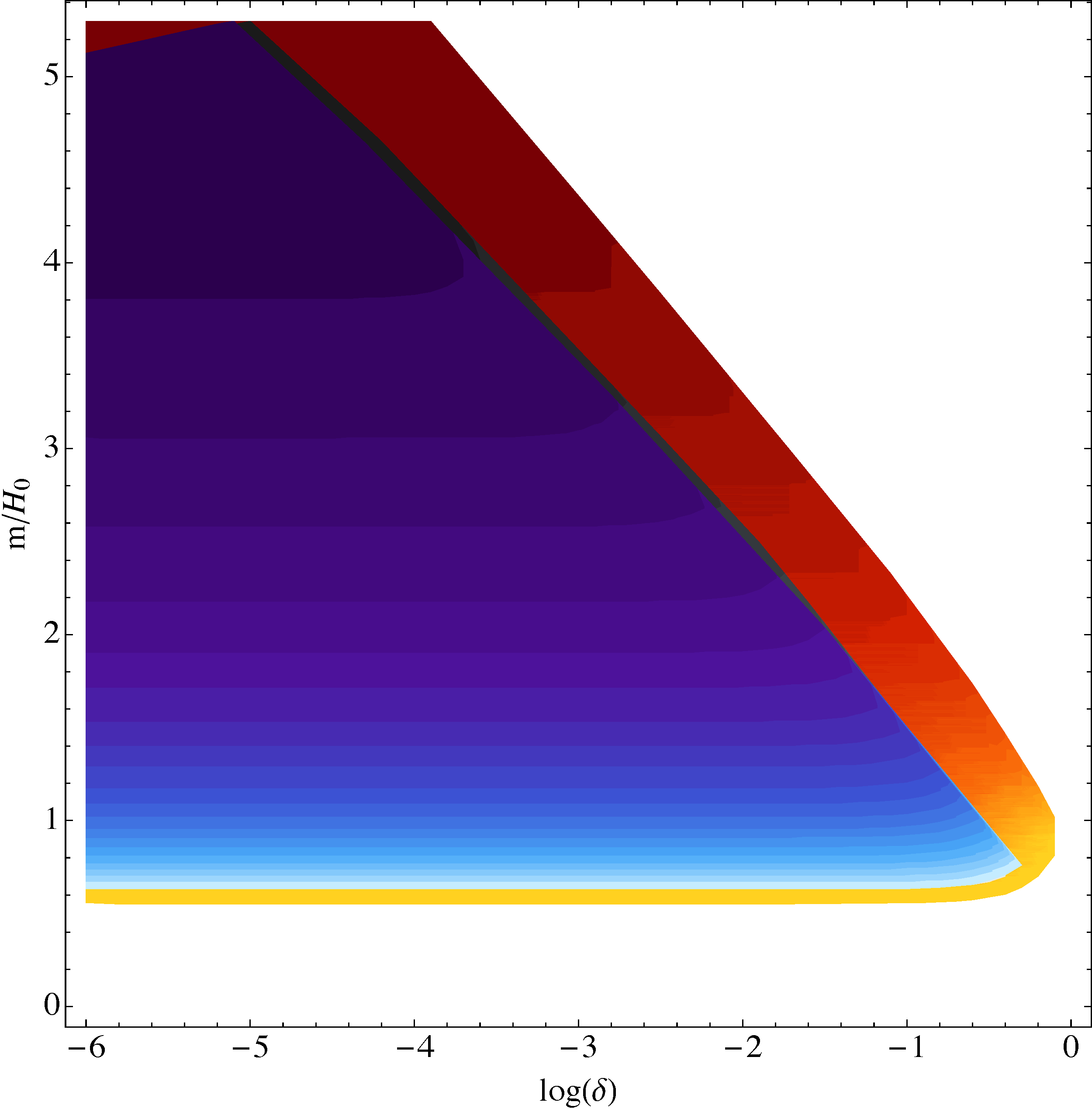}
\caption{Same as Fig.~\ref{fig:future_B} but for Model B.}
\label{fig:future_B}
\end{figure*}For Model A, we show the allowed region of $\{\delta, m,\alpha\}$ in Fig.~\ref{fig:D_A-H_z057-CMB_bossfid}, as constrained by BOSS (top left panel) and Planck (bottom left panel) measurements of $\{d_A(z), H(z)\}$ at $z=0.57$ and $\theta_{*}$ respectively. Predicted constraints (under the null hypothesis) from the SPHEREx satellite \cite{Dore:2014cca} are shown in the top right panel. The bottom right panel illustrates a comparison between the allowed regions from BOSS, Planck, and SPHEREx (predicted) measurements. We perform the same analysis for Model B, with results shown in Fig. \ref{fig:D_A-H_z057-CMB_bossfid_B}. 

The main difference between Models A and B is that at the same mass and Peccei-Quinn scale, Model B is considerably more finely tuned in $\delta$. This trend subsists in the forecasted constraints from future data, discussed below (for the null hypothesis).
We can see that there is a common general behavior, that as $\alpha$ decreases, the allowed mass increases and the allowed range of misalignment angles decreases. For sub-Planckian Peccei-Quinn scale $\alpha \lesssim 1$, the initial misalignment angle must be finely tuned to lie near the top of the quintessence potential, $\theta=\pi-\delta\simeq \pi$, and the axion mass $m\gtrsim \mathcal{O}(1)H_{0}$. 

Qualitatively, these results arise from the requirement that the quintessence field have sufficient energy density to explain current cosmic acceleration, without spoiling the agreement of the standard cosmology at higher redshifts, where nonrelativistic matter must dominate the cosmic expansion history. The area of the parameter space that is allowed from observational constraints is very similar when comparing current BAO and CMB measurements. Increasing the precision in BAO measurements shrinks the projection of the allowed region onto two dimensions, but not dramatically. 

We then investigated if having measurements of the BAO peak at higher redshift can help reduce the area of the allowed region. We predicted values of $\{d_A(z), H(z)\}$ at $z=1$ and $z=2$ using our full grid of axion models, and then compared with a Fisher forecast of the error bars of future galaxy surveys produced by the SPHEREx satellite, the Square Kilometre Array (SKA) (for a specific SKA-oriented BAO measurement prediction, see Ref.~\cite{Bull:2015nra}), and the WFIRST satellite mission. We predict the expected $95\%$~confidence constraints to the $\{\delta,m,\alpha\}$ parameter set. The results are shown in Fig.~\ref{fig:future} for Model A, and in Fig. \ref{fig:future_B} for Model B.

\begin{figure*}[tb]
\includegraphics[width=0.9\textwidth]{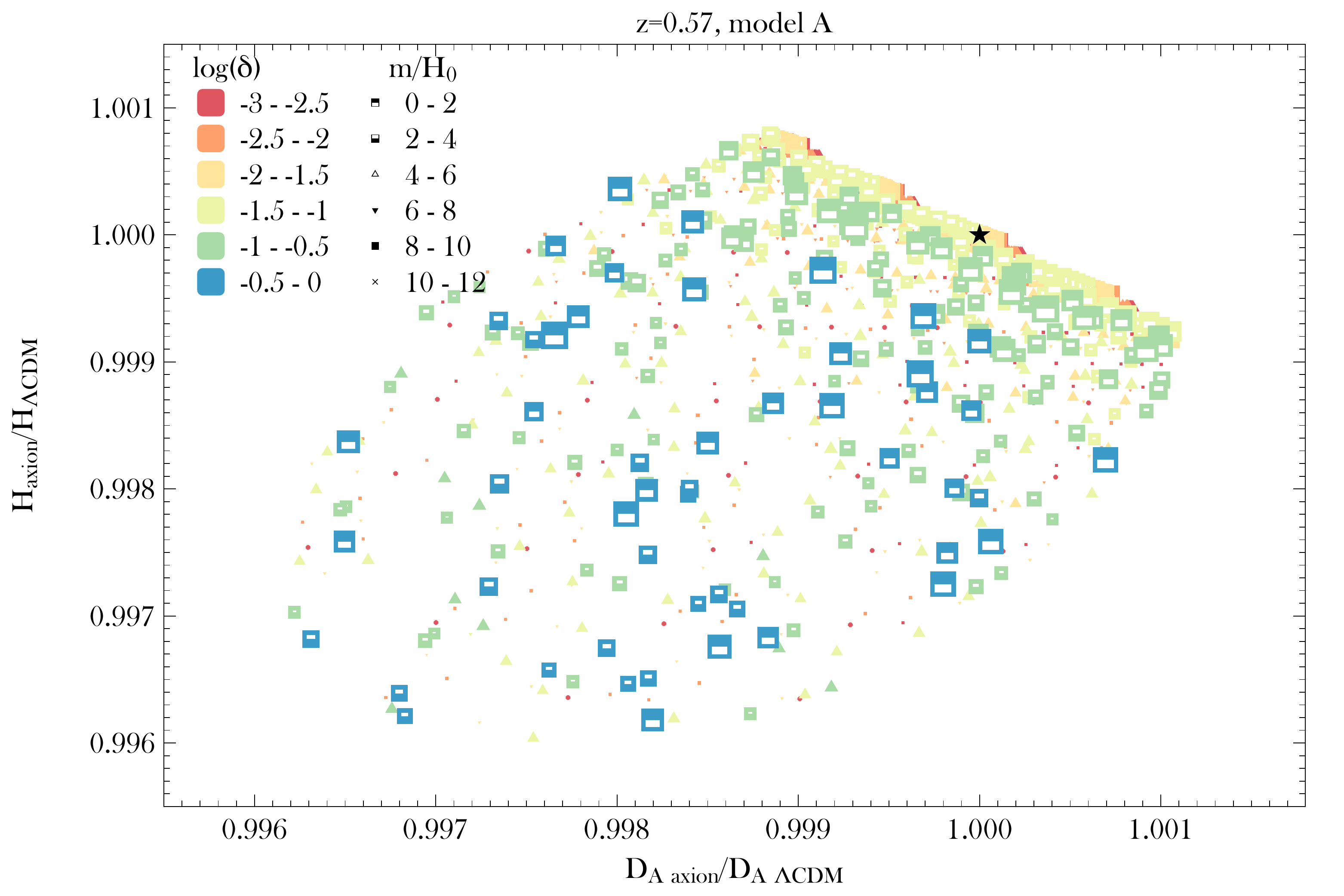}\\ 
\includegraphics[width=0.9\textwidth]{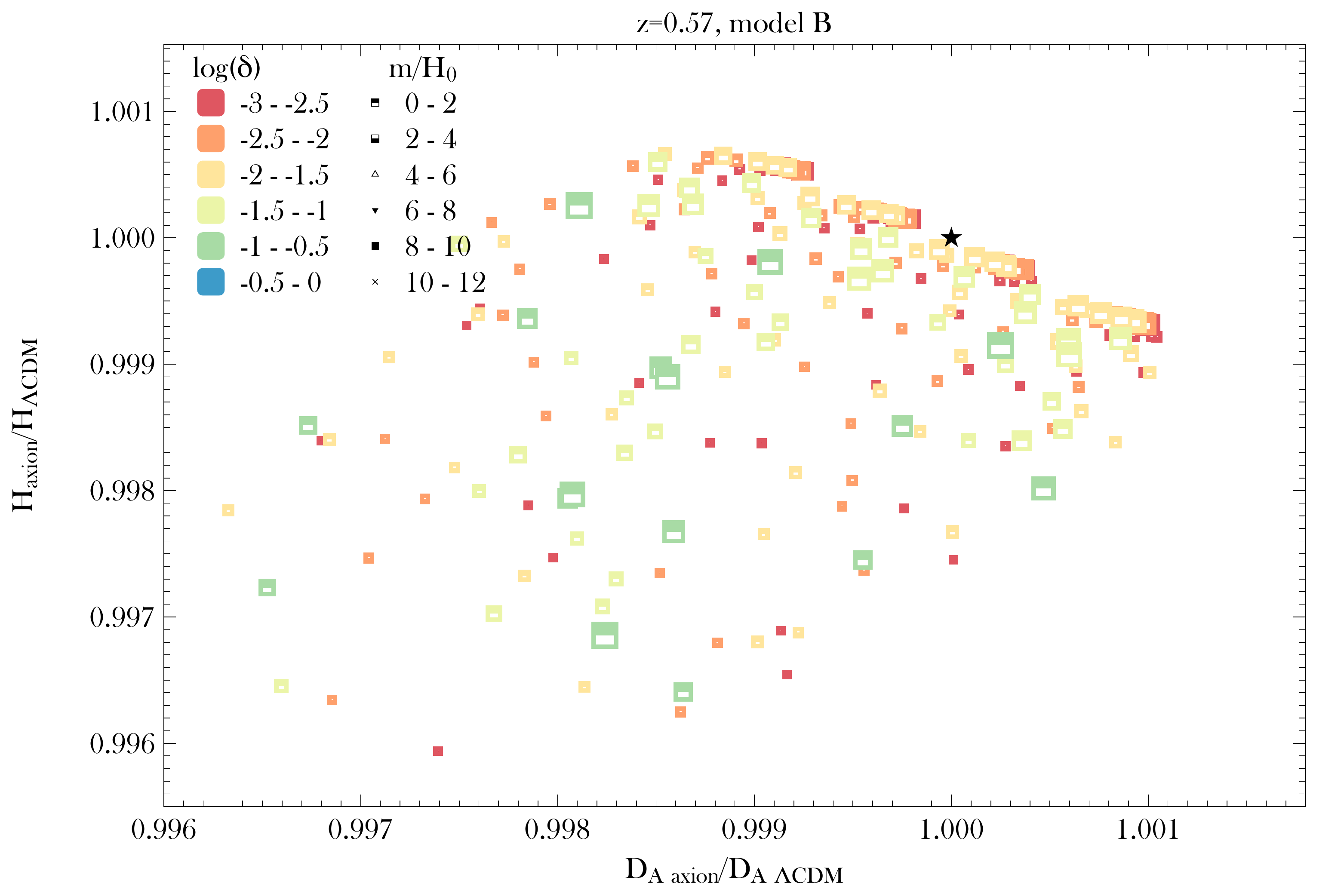}
\caption{Ratio of predicted $\{d_A, H\}$ to fiducial $\Lambda$CDM values for different combinations of the axion model parameters $\{\delta, m, \alpha\}$, for Model A (top panel) and Model B (bottom panel). Different values of $\delta$ are shown in colors, $m$ in symbol type and the value of $\alpha$ is proportional to the size of the symbol. The large star in both panels shows the case of $\Lambda$CDM fiducial values at current best-fit values from Planck.
}
\label{fig:ellipse}
\end{figure*}

Qualitatively, we see that if future experiments are consistent with the null hypothesis, the allowed region (as parameterized by the angle $\delta_{i}$) will become increasingly finely tuned. On the other hand, future experiments will not be \textit{dramatically} more sensitive to any $2$ of $\{\delta, m, \alpha\}$ than current measurements, in spite of the increasing precision in $\{d_A(z), H(z)\}$ measurements and the dramatic decrease in allowed parameter-space \textit{volume}. We can see why this happens in Fig.~\ref{fig:ellipse}, where we show the $95\%$~confidence level ellipse (in $d_A$ and $H$, divided by their fiducial values), for a futuristic experiment probing $d_A$ and $H$ at $0.1\%$ precision. 

The points in Fig. \ref{fig:ellipse} show predictions for different choices of parameters for the axion model. For each point, the color indicates the value of $\delta$, the symbol indicates the value of $m$, while the size of the symbol is proportional to the value of $\alpha$. We see that it is possible to get very close to the central values of $d_{A}(z=0.57)/d_{A}^{\rm fid}(z=0.57)\simeq 1.00$ and $H(z=0.57)/H_{\rm fid}(z=0.57)\simeq 1.00$ and still find allowed models with no dramatic change in the $(\delta,m)$ range consistent with the data. Here ${\rm fid}$ denotes values computed at fiducial (Planck) $\Lambda$CDM parameter values. An investigation of other possible observables affected by axiverse models is left for future work. So far, we have treated CMB probes of axiverse-inspired quintessence models using a simple constraint from the angular acoustic horizon $\theta_{*}$. In Appendix \ref{sec:cmb}, we verify that this simple treatment is equivalent to the result that would have been obtained through a full computation of CMB anisotropies in axiverse-inspired quintessence models.

\section{Conclusions}
\label{sec:conclusions}
By positing the existence of many light fields, the axiverse scenario raises the possibility that today's epoch of accelerated cosmic expansion results from a moderate fine tuning, in which today's acceleration occurs in a fraction $\sim 10^{-3}\to 10^{-2}$ of cosmological draws of the axion misalignment angle $\delta$, as long as the dimensionless Peccei-Quinn scale $\alpha=f_{\rm a}/M_{\rm pl}\gtrsim 0.1$. This is a vast improvement with respect to the required fine tuning of  quintessence models, which do not generically address the ``why now?" problem (``Tracker" models offer some improvement by eliminating the need for finely tuned initial conditions \cite{Peebles:1987ek,Ratra:1987rm,Caldwell:1997ii}, but they still require a finely tuned overall quintessence energy scale \cite{Peebles:1987ek,Ratra:1987rm,Caldwell:1997ii}).

The axiverse scenario motivates a family of quintessence models in which the field displacement from maxima of anharmonic potentials must be relatively small. After discussing the dynamics of these models, we showed that the dark energy equation-of-state, Hubble parameter at a variety of redshifts, angular-diameter distances, and CMB angular sound horizon deviate sufficiently from their fiducial values to allow precision cosmological probes of axiverse-inspired quintessence. 

We proceed to test these scenarios using two very different measurements of the baryon-photon sound horizon, the BAO feature measured in galaxy surveys and the angular scale of the CMB acoustic peaks. In terms of parameter space volume, the CMB is dramatically more constraining, eliminating a fraction $\sim 0.8$ of models that pass a well motivated dark energy prior. In the projected parameter space of $m$ and $\delta$ (for sub-Planckian $f_{\rm a}$), we find that the CMB and BAO measurements constrain axiverse-inspired quintessence models to comparable \textit{areas} in parameter space. In the future, cosmic-variance limited measurements of CMB polarization and very sensitive BAO surveys will be dramatically more sensitive, empirically distinguishing (from the $\Lambda$CDM model) all but $\sim 10^{-5}$ of the parameter space allowed by dark energy priors.

The constraints presented in this work were derived simply, varying the quintessence parameters of interest without performing a full cosmological Monte Carlo Markov Chain. In future work, we will pursue such techniques to explicitly account for priors on all parameters, and consider a variety of other observables, such as the galaxy-CMB cross-correlation (which may be more sensitive to the quintessence-induced ISW effect than the CMB temperature power spectrum), measurements of the cosmic expansion history from Type IA supernovae, direct kinematic tests of cosmic acceleration, and gravitational lensing (of the CMB and background galaxies).

\begin{acknowledgments}
This work was supported at JHU by NSF Grant No.\
0244990, NASA NNX15AB18G, the John Templeton Foundation, and the
Simons Foundation. R..E acknowledges the support of the New College Oxford-Johns Hopkins Centre for Cosmological Studies as well as a grant from Research Grants Council of the Hong Kong Special Administrative Region, China (HKUST4/CRF/13G). D.G. is funded at the University of Chicago by a National Science Foundation Astronomy and Astrophysics Postdoctoral Fellowship under Award No. AST-1302856. This work was supported in part by the Kavli Institute for Cosmological Physics at the University of Chicago through Frant NSF PHY-1125897 and an endowment from the Kavli Foundation and its founder Fred Kavli. J.P. is supported by the New Frontiers program of the Austrian Academy of Sciences. 
\end{acknowledgments}
\begin{appendix}
\section{Axiverse-inspired quintessence and the CMB}
\label{sec:cmb}
Dark energy contributes to the expansion history of the Universe but clusters very weakly (and indeed, is completely smooth in the limit of a pure cosmological constant). As a result, gravitational potential wells decay on scales that enter the horizon during dark energy domination. This late-time Integrated Sachs-Wolfe (ISW) effect imprints additional temperature anisotropies on the CMB at low multipole index $\ell$. The detailed evolution of $w$ in axion models will thus affect the precise shape of both the CMB temperature anisotropy power spectrum $C_{\ell}^{\rm TT}$ and galaxy-CMB temperature cross power spectrum $C_{\ell}^{T \rm g}$ \cite{Frieman:1995pm,Coble:1996te} at low $\ell$. Although the statistical significance of measurements of these quantities at low $\ell$ is limited by cosmic variance, there could be axion dark-energy models allowed by the constraints of Sec. \ref{sec:currentconstraints} that are further constrained by the imprint of the late-time ISW effect on $C_{\ell}^{\rm TT}$ and $C_{\ell}^{\rm T \rm g}$.

The quintessence-induced change to the angular sound horizon $\theta_{*}$ drives the CMB constraints/forecasts of Sec. \ref{sec:currentconstraints}, but here we explore if this simple geometric constraint captures the full power of the CMB to probe axiverse inspired quintessence models. We do this by computing in greater detail the effect of axiverse-inspired quintessence on CMB anisotropies.

The quintessence field $\phi$ may depend on spatial location as well as time, and must be permitted to carry spatial perturbations in a self-consistent computation of 
observable quantities like $C_{\ell}^{\rm TT}$ for a quintessence cosmology. The relevant equation of motion (the perturbed Klein-Gordon equation) can be rewritten as a pair of fluid equations in synchronous gauge \cite{Hu:1998kj,Bean:2003fb,Weller:2003hw,Hu:2004xd}, to ease their inclusion in the cosmological Boltzmann code \textsc{camb} \cite{cambnotes}:
\begin{align}
\delta'_{\phi}=&-ku_{\phi}-\left[1+w(a)\right]\dot{h}/2-3\mathcal{H}\left[1-w(a)\right]\delta_{\phi}\nonumber \\-&9\mathcal{H}^{2}\left(1-c_{\rm ad}^{2}\right)u_{\phi}/k,\label{eq:eoma}\\
u'_{\phi}=&~2\mathcal{H}u_{\phi}+k \delta_{\phi}+3\mathcal{H}\left(w-c_{\rm ad}^{2}\right)u_{\phi}\label{eq:eomb},\end{align}where $\delta_{\phi}$ is the fractional axion energy density perturbation, and the adiabatic sound speed is
\begin{equation}
c_{\rm{ad}}^2\equiv \frac{{P}_{\phi}^{'}}{{\rho}^{'}_{\phi}}= w-\frac{{w'}}{3\mathcal{H}\left(1+w\right)}\label{eq:adiabat_cs}.
\end{equation}

The dimensionless axion heat flux is $u_{\phi}=(1+w)v_{\phi}$ where $v_{\phi}$ is the scalar axion velocity perturbation. Gravitational potentials enter through the synchronous gauge metric perturbation $h$. The conformal Hubble parameter is given by $\mathcal{H}=a'/a=aH$, where $'$ here denotes a derivative with respect to conformal time $\eta$, defined by $d\eta=dt/a$.

The gravitational imprint of axions is in turn included through the synchronous gauge Einstein equations, with axionic pressure, energy density, and momentum flux contributions
\begin{eqnarray}
\delta P_{\phi}&=&\rho_{\phi}\left[\delta_{\phi}+3\mathcal{H}(1-c_{\rm ad}^{2})v_{\phi}/k\right]\label{dpdef},\\
\delta \rho_{\phi}&=&\rho_{\phi}\delta_{\phi},\\
\left(\rho_\phi+P_{\phi}\right)v_{\phi}&=&\rho_{\phi}u_{\phi}.\label{eq:dqdef}
\end{eqnarray}In recent work by one of us and other collaborators \cite{Hlozek:2014lca}, these equations were self-consistently implemented in the CMB Boltzmann code \textsc{camb} \cite{cambnotes} for a quadratic potential (in that case, allowing for both slowly rolling axion dark energy and coherently oscillating axion dark matter), along with the accompanying equations for the mean axion energy-density and pressure, which affect the recombination and decoupling history of the universe. 

Here, we have generalized this modified \textsc{camb} \cite{cambnotes} using the full Model A and Model B anharmonic potentials in Eqs.~(\ref{eqn:standardpotential})-(\ref{eqn:alteredpotential}), but restrict ourself to the ``dark-energy like" part of parameter space in the language of Ref. \cite{Hlozek:2014lca}. We solve the quintessence perturbation equations of motion numerically. The axion potential enters only through the evolution of the equation-of-state $w(a)$ and axion adiabatic sound speed $c_{\rm ad}$. We thus solve Eq.~(\ref{eqn:EOM}) numerically, evaluating $w(a)$ and $c_{\rm ad}^{2}$ as described in detail in Ref. \cite{Hlozek:2014lca}. These functions are then tabulated and interpolated by our modified version of \textsc{camb} to obtain cosmological observables like $C_{\ell}^{\rm TT}$. Unlike the treatment in Ref. \cite{Hlozek:2014lca}, we do not solve Eqs.~(\ref{eq:eoma})-(\ref{eq:eomb}) with a WKB approximation. 

\begin{figure*}[tb]
\includegraphics[width=0.7\textwidth]{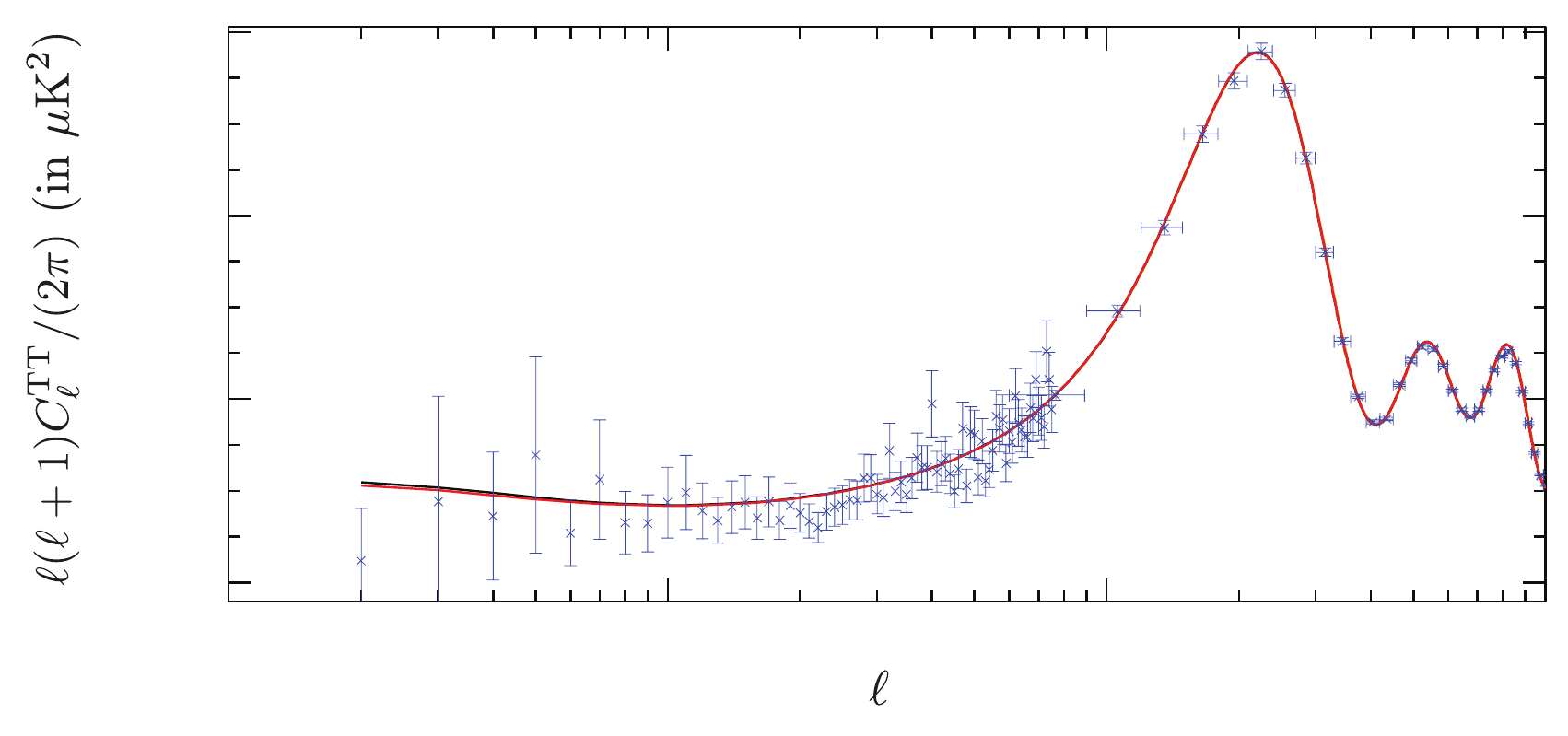}
\caption{CMB temperature anisotropy power spectra $C_{\ell}^{\rm TT}$ for an axiverse-inspired quintessence model with the largest deviation from $\Lambda$CDM in the grid of Sec. \ref{sec:currentconstraints}, generated using a modified version of \textsc{CAMB}. This power spectrum is then remapped in angular scale via $\ell \to \ell \times \theta_{*}/\theta_{*}^{\rm fid}$ to highlight the late-time ISW effect. The quintessence case is shown in red, while the $\Lambda$CDM case is shown in black. Blue points with error bars are published Planck band powers from Ref. \cite{Ade:2015xua}.}
\label{fig:cltt}
\end{figure*}

For both Models A and B, we generate 790 uniformly sampled (in $\log{\delta}$, $m$, and $\log{\alpha}$) subsets of the full grid of models used to generate constraints in Sec. \ref{sec:currentconstraints} and generate the resulting CMB power spectra. One example (for Model A) is shown in Fig. \ref{fig:cltt}, which shows the $\Lambda$CDM $C_{\ell}^{\rm TT}$ power spectrum, along with the results of the full quintessence-modified \textsc{camb} computation, after applying the mapping $\ell \to \ell \times \theta_{*}/\theta_{*}^{\rm fid}$, where $\theta_{*}^{\rm fid}$ is the angular sound horizon in our fiducial cosmology (with no axiverse-inspired quintessence). This mapping should undo the change to $C_{\ell}^{\rm TT}$ induced by quintessence-induced changes to the angular sound horizon and isolate the effect induced by the late-time ISW effect. 

\begin{figure*}[tb]
\includegraphics[width=\textwidth]{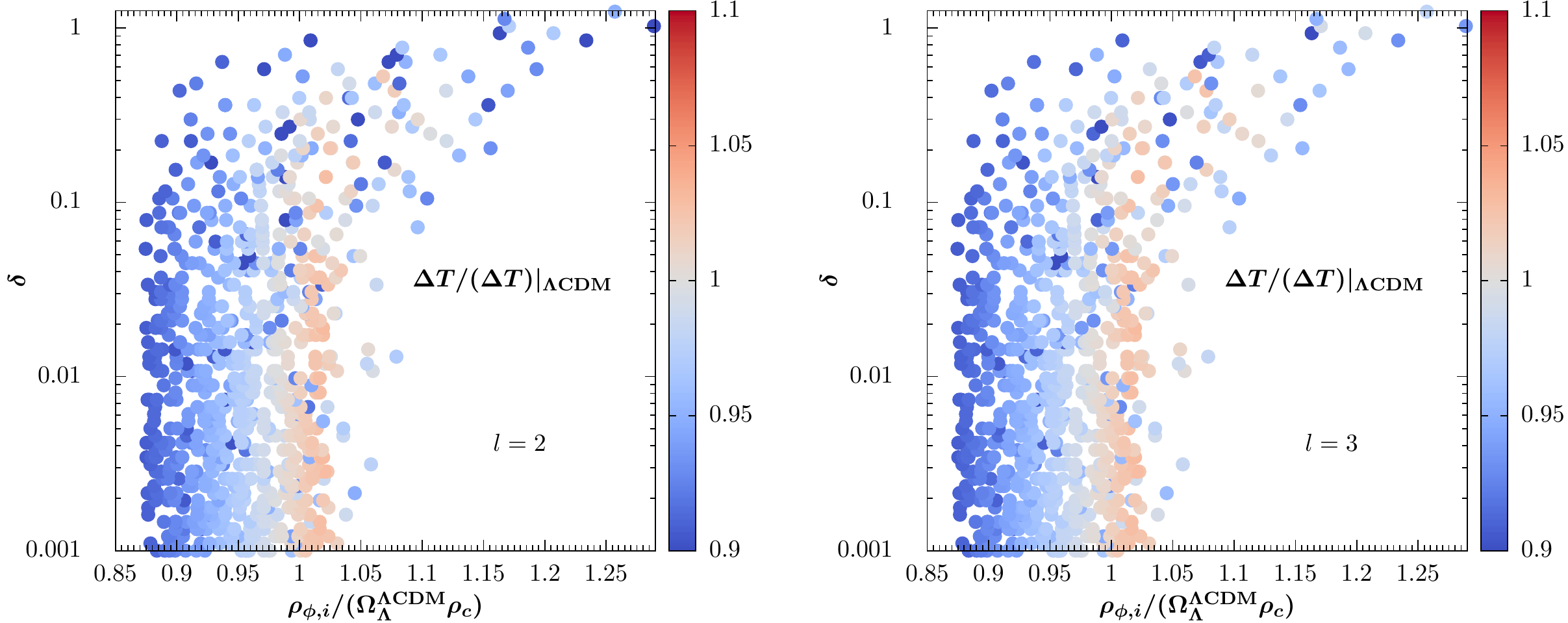}
\caption{For Model A, the ratio of ISW-induced CMB temperature anisotropies comparing axiverse-inspired quintessence models with the fiducial $\Lambda$CDM model for modes corresponding roughly to $\ell=2$ and $\ell=3$ temperature multipole moments as a function of dark energy density and initial misalignment angle $\delta$. A prior of $w\leq -0.7$ and $0.6\leq \Omega_{\phi}\leq 0.7$ was imposed to subselect from the grid of $790$ models uniformly sampled in $\log{\delta}$, $m$, and $\log{\alpha}$.}
\label{fig:isw}
\end{figure*}

We see that $C_{\ell}^{\rm TT}$ looks almost indistinguishable from the $\Lambda$CDM case after remapping the sound horizon, with some visibly discernible deviation from the fiducial case for $\ell \leq 50$. We find that the fractional change to $C_{\ell}^{\rm TT}$ is negligible, compared both with cosmic variance and the benchmark for noise-free parameter fitting, $|\Delta C_{\ell}^{\rm TT}/C_{\ell}^{\rm TT}|=3/\ell$ \cite{Seljak:2003th}. We can see that even for this model (which shows the most dramatic deviation from $\Lambda$CDM in our subsample after remapping the angular sound horizon) that the modifications to the late-time ISW effect induced by the quintessence field are well below cosmic variance at all scales and hence undetectable. In other words, the constraints imposed in Sec. \ref{sec:currentconstraints} using CMB measurements of $\theta_{*}$ are robust for axiverse-inspired quintessence models.

We find that remapping of $\theta_{*}$ is able to absorb $\sim 99\%$ of the change in $\chi^{2}/{\rm DOF}$ between the quintessence and $\Lambda$CDM model, showing that additional changes to $C_{\ell}^{\rm TT}$ from the late-time ISW effect are negligible. The much larger change in the ISW effect between axion and $\Lambda$CDM models seen in
Ref. \cite{Hlozek:2014lca} results from the much wider range of axion masses considered in that work, where axions were considered as both dark matter \textit{and} dark energy candidates. 

As \textsc{camb} includes many different physical effects, not just the ISW effect, it is interesting to consider a simpler calculation that isolates the effect of dark energy on the ISW effect. To this end, we implement the equations of Ref. \cite{PhysRevD.82.083522} for the ISW effect sourced by cosmological perturbation modes with a variety of comoving wave numbers $k$.  We assume an arbitrary initial perturbation normalization (with no additional power spectrum induced scale dependence). We use the same $790$ models as above but impose the ``dark energy" prior discussed above. 

We then compare the results to the null hypothesis, with results shown in Fig. \ref{fig:isw} for modes roughly corresponding to the multipoles $\ell=2$ and $\ell=3$. The correspondence between wave number is treated approximately, via $\ell \simeq \pi/\theta=\pi d_{a}(z_{\rm rec})/r_{\rm dec}$. We see from Fig. \ref{fig:isw} that the $\sim 5\%$ change in CMB anisotropies at low $\ell$ is much smaller than cosmic variance, confirming our conclusion that the late-time ISW induced change to CMB anisotropies from axiverse-inspired quintessence is undetectable.\footnote{Values for the ISW-only computation differ from the computation using \textsc{camb}, due to the fact that other terms in the line of sight solution were neglected, and the fact that no integral over the power spectrum was done here.}

\end{appendix}

%\bibliography{bibliography}

\end{document}